\documentclass[floatfix, preprintnumbers, amsmath, amssymb, 10pt]{revtex4}
\usepackage{epsfig}
\usepackage{dcolumn}
\usepackage{bm}

\begin{document}

\date{\today}

\title{Moduli (Dilaton, Volume and Shape) Stabilization via Massless F and D String Modes}

\author{Subodh P. Patil $^{1,2,3,4)}$} \email[email: ]{patil@het.brown.edu}
\affiliation{1) Dept.of Physics, Brown University, 
Providence R.I. 02912, U.S.A.} 
\affiliation{2) Dept.of Physics, Mc Gill University, Montr\'eal QC, H3A 2T8, Canada}
\affiliation{3) Theoretical Physics, Queen Mary University of London, London, E1 4NS, U.K.}
\affiliation{4) DAMTP, Cambridge University, Cambridge CB2 0WA, U.K.}

\begin{abstract}
Finding a consistent way to stabilize the various moduli fields which generically appear in string theory compactifications, is essential if string theory is to make contact with the physics we see around us.   
We present, in this paper, a mechanism to stabilize the dilaton within a framework that has already proven itself capable of stabilizing the volume and shape moduli of extra dimensions, namely string gas cosmology. Building on previous work, which uncovered the special role played by massless F-string modes in stabilizing extra dimensions once the dilaton has stabilized, we find that the string gas cosmology of such modes also offers a consistent mechanism to stabilize the dilaton itself, given the stabilization of the extra dimensions. We then generalize the model to include D-string gases, and find that in the case of bosonic string theory, it is possible to simultaneously stabilize all the moduli we consider consistent with weak coupling. We find that our stabilization mechanism is robust, phenomenologically consistent and evades certain difficulties which might previously have seemed to generically plague moduli stabilization mechanisms, without the need for any fine tuning.    
\end{abstract}

\newcommand{\eq}[2]{\begin{equation}\label{#1}{#2}\end{equation}}

\maketitle

\section{Introduction}

The question of how to stabilize the multitude of moduli fields that arise in any attempt at obtaining 4-d effective physics from string theory, is at present, still an open one. Although there have been many notable attempts in the last few years (see \cite{eva} and references therein), many issues remain open. The majority of these attempts are crouched in the framework of low energy effective field theory, where either ad-hoc moduli potentials, potentials generated by higher order (e.g. in $\alpha'$) effects, or potentials generated by fluxes are used to stabilize the various moduli fields. In spite of the various successes encountered by these attempts, they come up against some quite general problems, consistency with late time cosmology not least among them. A specific problem that accompanies moduli stabilization schemes that use potentials to fix the moduli fields, is the fact that any such potential will generate a cosmological constant at its minima, which introduces a host of other theoretical problems \cite{SG}.
\par

String gas cosmology (SCG) \cite{BV}\cite{TV}\cite{kp}\footnote{See \cite{sa2}\cite{sw2}\cite{sw3}\cite{sp1}\cite{sp2}\cite{edna}\cite{nyc1}\cite{SGC} for other work on SGC.}, on the other hand, has to date focused on a narrow subset of the moduli problem-- that of stabilizing extra dimensions. From its origins as a mechanism in which a 3+1 dimensional universe is generated by the dynamics of strings in the early universe \cite{BV}\cite{TV}, it has become clear that the effects of a string gas can stabilize any number of extra dimensions at the string scale in a way that yields a late time scenario that corresponds to our present day universe \cite{sp1}\cite{sp2}, provided that the dilaton has been fixed through some external mechanism. Key to this stabilization mechanism, and to the consistency of the SCG program in general, are the properties of a gas of massless string modes. It is the goal of this paper to report on even further-reaching consequences of massless string gas cosmology, which evidently contains a natural mechanism to stabilize all moduli fields for a particular toy model (bosonic string theory compactified on a torus).   

\par

We will shortly provide an a priori motivation for focusing specifically on massless string modes if SGC hopes to be theoretically and phenomenologically consistent, after which we will begin in earnest our attempt at moduli stabilization using these modes. Before we do any of this however, we should offer a brief (and neccesarily incomplete) historical review of the SCG program, in order to give us perspective on why we feel it is rather strategically placed to solve the moduli problem. 

\par

The Brandenberger-Vafa (BV) mechanism (proposed in \cite{BV}), posits that the observed dimensionality of macroscopic spacetime arises from the dynamics of strings in the early universe. Such a mechanism was put on firmer footing in \cite{TV}, where it was reasoned that dilaton gravity (i.e. a dynamical dilaton) was neccesary for such a mechanism to work. Such reasoning was based on the observation that any positive energy density and negative pressure (which accompany winding string states) in gereral cause the expansion of spacetime in general relativity, rather than its contraction or stabilization, whereas for dilaton gravity, negative pressures generically cause contraction. In the years that followed, much work has gone in to putting the ideas of \cite{BV} on a firmer footing. Among the more notable results has been the demonstration that the BV mechanism still holds once we include branes of various dimensions as well as strings as the fundamental degrees of freedom \cite{sa2}. Since then, it has been shown that, in the context of dilaton gravity, string gases can effectively stabilize extra dimensions at the string scale \cite{sw2}, with a resulting isotropy of the non-compact dimensions \cite{sw3}. More recently, it has been shown that string gases can stabilize any number of extra dimensions even after the dilaton has stabilized \cite{sp1}\cite{sp2}, in a way that yields a resulting late time FRW cosmology, and which has a consistent phenomenology (which we will uncover for oursleves further on). There is beginning to emerge, a surprisingly complete cosmology that results from having string gases pervade the universe, which we hope will eventually provide us with testable predictions on the nature of dark matter and on effects on the microwave background \cite{sp3}, in addition to providing a framework in which one can potentially model non-singular bouncing cosmologies \cite{sp4}. Even more recently, it has been shown \cite{edna} that, in addition to stabilizing the moduli corresponding to the radii of the various cycles of a toroidal compactification, the shape moduli are also stabilized. Emboldened by this, it might appear that if SGC can stabilize the dilaton itself, whilst preserving the stabilization of all of the moduli associated with the compactified dimensions, then it might serve as an example of how moduli stabilization in more complicated string theory settings might be effected. 

\par

As mentioned previously, the workhorse for this paper is going to be a gas of massless string modes. Exactly why this should be, will need no further justification if we can demonstrate moduli stabilization using these modes, but attempting an a priori answer to this question provides a useful check on the theoretical and phenomenological consistency of what we are doing, and answers some of the criticisms that have previously been raised against SGC. 

\subsection{The Case For Massless String Modes} 

Consider first the idea that a gas of strings can stablize the various moduli fields that might appear in our particular string compactification. An immediate objection one might have is that there is no way to do this without severely overclosing the universe. That is, since all string states have masses which go like the Planck mass (if we equate this to the scale of stringy physics):

\eq{m2}{m \sim \frac{N}{\sqrt{\alpha'}} \sim M_{pl}N,}

\noindent where $N$ is a number of order unity for the lowest lying string states, and $M_{pl} = 2.43 \times 10^{18} GeV$ is the 4-d Planck mass, one might expect a fluid of such strings to be phenomenologically ruled out. However, if we consider a gas of such strings which are massless, then one has a shot of evading phenomenological bounds. In \cite{sp1} and \cite{sp2} it was shown that a gas of massless string modes can effectively stabilize any number of extra dimensions at the string scale without overclosing the universe, whilst simultaneously providing a strong enough stablization mechanism that evades fifth-force constraints \footnote{Recall that from the perspective of the non-compact directions, fluctuations in the radion fields look like a scalar field with a mass$^2$ given by the curvature of the potential for the radion. It turns out that this corresponding mass can be quite easily made large enough to avoid introducing any fifth-force interactions \cite{sp1}\cite{sp2}.}. However, there is another important theoretical reason for us to focus exclusively on massless string modes. We will very shortly be considering dilaton gravity in the presence of string sources, and accordingly will be in the regime of a low energy effective theory described by the action:

\eq{dact}{S = \frac{1}{2\kappa_0^2}\int d^{D+1}x\sqrt{-g} e^{-2\Phi}\Bigr(R + 4\partial_\mu\Phi\partial^\mu\Phi - \frac{1}{12}H_{\mu\nu\lambda}H^{\mu\nu\lambda}\Bigr).}  
      
\noindent Where $\Phi$ is the dilaton, $g_{\mu\nu}$ is the $D+1$ dimensional metric tensor, $H_{\mu\nu\lambda}$ the field strength of the two form field $B_{\mu\nu}$, and $\kappa_0^2$ is the dimensionful normalization of this action. This is specifically the `massless background' \cite{pol}\cite{TV} of the low energy effective fields, as it is only valid for curvature regimes such that:

\eq{cr}{\alpha'R_c \ll 1,} 

\noindent where $R_c$ is the characteristic curvature of the background in question, and $\alpha'$ is related to the string tension as $T = 1/2\pi\alpha'$. As long as such a condition is satisfied, the curvature scale of the background is much bigger than the string scale, and massive string states will not be created. Conversely, if we wanted to introduce massive string states, then we would have to include higher order corrections in $\alpha'$. Hence if we are to introduce a gas of strings as the matter content for the action above, then it should only consist of massless string states, as this action describes a consistent background for these states alone.

\par

A potential concern that one might have however is that, naively, one might violate (\ref{cr}) if we are considering compactifications of the extra dimensions that are at the string scale. However, it turns out that when we consider the spacetime dynamics of such a gas on a toroidally compactified background, which eventually results in the stabilization of the extra dimensions and the dilaton, and compute the components of the Ricci scalar throughout the evolution of the system, we will find that the condition (\ref{cr}) is satisfied to an excellent degree. Hence our framework is consistent throughout, and we will be justified in using the low energy effective action (i.e. dilaton gravity) as the framework for our physics. Further issues concerning aspects of the string spectrum which we will be using are considered in the appendix, where we derive the energy momentum tensor for a gas of strings the backgrounds of interest for this report. There, we find that a string gas contributes an energy momentum tensor that can be derived from the following action:

\eq{mod}{S = -\int d^{D+1}x \sqrt{g_{00}} E[g_s,\phi],}

\noindent where $E$ is the internal energy of our string gas, which depends on the spatial part of the metric and the dilaton field. From this, we immediately infer that SGC has several compelling advantages over the effective field theory approach to moduli stabililzation. We see from the form of the above action, that one does not have the usual fine tuning problems associated with introducing potentials for the moduli fields. Such a potential would contribute to the energy momentum tensor from the action:

\eq{modbad}{S = -\int d^{D+1}x \sqrt{g} V(\phi),}      

\noindent where one would have to fine tune the minimum of the potential to vanish at the stabilization point to avoid an unacceptably large De-Sitter phase of evolution after stabilization. We see from (\ref{mod}) that the coupling of the string gas to gravity is rather more complicated than just the usual $\sqrt g \times V$ dependence, and the 'potentials' one derives from the moduli fields from (\ref{mod}) do not require their minima to be fine tuned at all, and in fact, quite generically result in Freidman-Roberston-Walker (FRW) expansion of the non-compact dimensions\cite{sp2}\footnote{Of course, if one wanted to incorporate a midly De-Sitter late time phase into the subsequent cosmology, then SGC at present has nothing to offer. However, it might be that the present state of acceleration is due to the effects of the string gas alone \cite{mersini}\cite{ras}}. In this way, we evade the negative results of Giddings \cite{SG} concerning radial moduli stabilization.

\par

However from a purely phenomenological point of view, SGC has yet another advantage over the field theoretical approach to moduli stabilization: consider the fact that any potential one introduces to stabilize the moduli will, from the prespective of the 4-d effective theory, cause the modulus in question to have a mass given by:

\eq{modmass}{m^2 \sim V''(\phi),}

\noindent whereas the contribution to the energy density of the universe of such a stablization potential will go like:

\eq{moded}{\rho \sim V(\phi),}

\noindent which, if we are close to the minimum of the potential, is also proportional to $m^2$. When we try to make contact with present day observations, the simultaneous requirement that the mass of the moduli field be large enough so that we evade fifth-force constraints, whilst small enough so as not to overclose the universe becomes very difficult to satisfy, and introduces its own host of fine-tuning problems. The case for string gases is markedly different. We will see further on that the 4-d masses of the moduli fields generated by the string gas go like:

\eq{sgmass}{m^2 \sim \frac{\mu_0}{Ea^3},}

\noindent where $E$ is the energy of a constituent massless string, $\mu_0$ is a number density and $a$ is the scale factor of the non-compact dimensions. This is to be compared with the contribution of the string gas to the energy density of the universe, which goes like:

\eq{sged}{\rho \sim \frac{\mu_0 E}{a^3}.}   

\noindent We notice that the two have an inverse dependence on the energy (i.e. center of mass momentum) of the massless strings in question. This is a situation which would be impossible to obtain from a field theoretical framework, and as we will see later, quite easily allows us to  evade fifth-force constraints whilst avoiding overclosing the universe without any fine tuning. Hence, we feel that SGC is a framework that is a priori, very well qualified to solve the moduli problem, the resolution of which we will turn to shortly.

\subsection{Outlook, Outline and Conventions}

Since we do not wish to prejudice the applicability of this work to a particular string theory (i.e. a particular corner of moduli space), we leave the dimensionality of spacetime, and the number of compact dimensions general, we only assume that some dimensions are toroidally compactified and that F-strings (and later, D-strings) are admitted as stable degrees of freedom. We will, of course, pick specific cases when the need arises. The outline of the paper is as follows: borrowing from results in the appendix, where we derive the energy momentum tensor of a gas of strings in our background (following \cite{sp1}), we then set up the model in section 2. After carefully considering the question of frame switching, we study the dynamics of the coupled string gas/dilaton gravity system, and review the result \cite{sp2} that, given a stable dilaton, the extra dimensions are stabilized at the string scale. We then show that we obtain dilaton stabilization given a background with stabilized extra dimensions. Next, we demonstrate that the coupled radion-dilaton system also stabilizes, but in a somewhat unsatisfactory manner, in that one of the modes of the coupled system becomes becomes massless from the perspective of the 4-d effective theory (and hence phenomenologically unacceptable). We then generalize our framework to include D-string degrees of freedom, and find that in certain cases, possibly involving the breaking of supersymmetry, we can simultaneously stabilize the dilaton and the moduli describing the extra dimensions.
\par
Our conventions are as follows: we work in $D+1$ spacetime dimensions,with the $D = d + p~$ spatial dimensions partitioned into $d$ non-compact dimensions and $p$ compact dimensions. We use the `mostly plus' metric with signature $(-,+,+,..,+)$, and work in natural units ($\hbar = c = 1$) with all dimensions expressed in terms of electron volts (eV). Spacetime indices will be denoted by Greek indices \{$\mu,\nu...$\}, with the non-compact spatial indices denoted by lower case Latin indices from the middle of the alphabet \{$i,j,k...$\}, and compact spatial indices, with lower case Latin indices from the beginning of the alphabet \{$a,b,c...$\}.      

\section{The Model}

Consider the dilaton gravity action in the string frame (where we set the form fields to be vanishing):

\eq{da}{S = \frac{1}{2\kappa_0^2}\int d^{D+1}x\sqrt{-g} e^{-2\Phi}(R + 4\partial_\mu\Phi\partial^\mu\Phi).} 

\noindent By performing the conformal transformation:

\eq{ct}{g_{\mu\nu} = e^{-2\omega}\tilde{g}_{\mu\nu},}

\noindent and using the relation (where tilded quantities refer to the fact that they involve $\tilde{g}_{\mu\nu}$):

\eq{R}{R = e^{2\omega}(\tilde{R} + 2D\tilde{\nabla}^2\omega - D(D-1)\tilde{\partial}_\mu\omega\tilde{\partial}^\mu\omega),}

\noindent we can re-express the above action (after integrating the term involving $\tilde{\nabla}^2$ by parts) as:

\eq{dae}{S = \frac{1}{2\kappa_0^2}\int d^{D+1}x\sqrt{-\tilde{g}} e^{-\omega(D-1) -2\Phi}\Bigl(\tilde{R} + D(D-1)\tilde{\partial}_\mu\omega\tilde{\partial}^\mu\omega + 4\tilde{\partial}_\mu\Phi\tilde{\partial}^\mu\Phi + 4D\tilde{\partial}_\mu\Phi\tilde{\partial}^\mu\omega\Bigr).}

\noindent Now, were we to set:

\eq{fix}{\omega = 2\frac{(\Phi_0 - \Phi)}{(D-1)},}

\noindent we could re-express the above action as:

\eq{se}{S = \frac{1}{2\kappa_0^2 e^{2\Phi_0}}\int d^{D+1}x\sqrt{-\tilde g} \Bigl(\tilde R - \frac{4}{(D-1)}\tilde\partial_\mu\Phi\tilde\partial^\mu\Phi\Bigr),}

\noindent which is the action for Einstein gravity coupled to a scalar field. We henceforth refer to the metric $\tilde{g}$ as the Einstein frame metric. We observe that the $D$ dimensional Newton's constant is given by:

\eq{newt}{8\pi G_D = \kappa_0^2e^{2\Phi_0}.}

We now introduce a gas of strings as the sources for the dilaton gravity action, which in its present form sees the dilaton as a scalar field minimally coupled to Einstein gravity. The energy momentum tensor for a gas of strings in one extra dimension was derived in \cite{sp1}, and generalized to any number of extra dimensions in \cite{sp2}. It was found there that proceeding from the Nambu-Goto action for a single string, and then hydrodynamically averaging, one would obtain the same energy momentum tensor for a string gas in the string frame if one were to derive this from the interaction term given by: 

\eq{int}{S_{int} = -\int d^{D+1}x \sqrt{-g_{00}}\mu_0\epsilon[g].}
 
\noindent Where $\epsilon$ is the energy of a single string (on which we will elaborate further on) in the given background, and $\mu_0$ is the number density of the string gas (in the string frame) with the metric dependence factored out:

\eq{nd}{\mu = \frac{\mu_0}{\sqrt{g_s}},}

\noindent where $g_s$ is the spatial part of the metric. Recall that once the dilaton has stabilized, the string frame and the Einstein frames are equivalent. Having taken dilaton stabilization for granted, it was shown in \cite{sp1} and \cite{sp2} that a phenomenologically consistent stabilization of the extra dimensions at the self dual radius follows, with a resulting FRW late-time cosmology. 

\par

Since it is the goal of this paper to show that, in addition to this, string gas cosmology is also capable of stabilizing the dilaton, we should consider the question of frame switching seriously. For this reason, and for completeness in general, we derive the energy momentum tensor of a gas of strings in the Einstein frame in the appendix. It turns out that after doing the analysis, one can  obtain the energy momentum tensor of a string gas in the Einstein frame by transforming the above interaction lagrangian using (\ref{ct}):

\eq{inte}{S_{int} = -\int d^{D+1}x \sqrt{-\tilde g_{00}}e^{-\omega}\mu_0\epsilon[\tilde g e^{-2\omega}],} 

\noindent where $\omega$ is given by (\ref{fix}). Defining the metric of the compact directions in the string frame as $\gamma$, we see that:

\eq{transf}{\gamma_{ab} = e^{-2\omega}\tilde{\gamma}_{ab},}

\noindent where $\tilde{\gamma}_{ab}$ is the metric for the compact directions in the Einstein frame. We use the result \cite{pol} that the energy of a closed string in the string frame, in the given background is given by: 

\eq{energy}{\epsilon = \sqrt{|p_{d}|^2 + (n,\gamma^{-1} n) + \frac{1}{\alpha'^2}(w,\gamma w) + \frac{1}{\alpha'}[2(n,w) + 4(N-1)]},}

\noindent where $p_d$ is the center of mass momentum along the $d$ non-compact directions and $(~,~)$ is a p-dimensional scalar product where $p$ is the number of compact directions. $n_a$ and $w^a$ are p-dimensional vectors respectively describing the momentum and winding quantum numbers along the $a^{th}$ compact direction, and $N$ is the string oscillator level. Note that in the above, we have eliminated the left moving oscillators $\tilde{N}$, using the level matching constraint \cite{sp1}\cite{sp2}:

\eq{lmc}{(n,w) + N = \tilde{N}.}

\noindent From this we find that (\ref{inte}) then becomes:

\eq{inte2}{S_{int} = -\int d^{D+1}x \sqrt{-\tilde g_{00}}\mu_0\sqrt{|\tilde{p}_{d}|^2 + (n,\tilde\gamma^{-1} n) + \frac{e^{-4\omega}}{\alpha'^2}(w,\tilde\gamma w) + \frac{e^{-2\omega}}{\alpha'}[2(n,w) + 4(N-1)]},} 

\noindent from which we define:

\eq{tenergy}{\tilde\epsilon = \sqrt{|\tilde{p}_{d}|^2 + (n,\tilde\gamma^{-1} n) + \frac{e^{-4\omega}}{\alpha'^2}(w,\tilde\gamma w) + \frac{e^{-2\omega}}{\alpha'}[2(n,w) + 4(N-1)]}.}

\noindent Deriving the energy momentum tensor as $\tilde{T}^\mu_\nu = \frac{2}{\sqrt{\tilde{g}}}\frac{\delta S}{\delta \tilde{g}_{\mu\lambda}}\tilde{g}_{\lambda\nu}$, we find that:

\begin{eqnarray}
\label{rho}
\rho &=& \frac{\mu_0\tilde\epsilon}{\sqrt{\tilde{g}_s}},\\
\label{pi}
p^i &=& \frac{\mu_0}{\sqrt{\tilde{g}_s}\tilde\epsilon}|p_{d}|^2/d,\\
\label{pa}
p^a &=& \frac{\mu_0}{\sqrt{\tilde{g}_s}\tilde\epsilon}[\frac{n_a^2}{\tilde{b}_a^2} - \frac{w_a^2 \tilde{b}_a^2}{\tilde\alpha'^2}].
\end{eqnarray}

\noindent From now on, for simplicity we will drop the tildes on the metric quantites in the Einstein frame, in which we remain from now on. The metric tensor in the Einstein frame is taken to be $diag[-1,a(t),a(t),a(t),b(t)...b(t)]$, $d$ being the number of non-compact directions, and $\tilde\alpha'$ given by:
 
\eq{redef}{\tilde{\alpha}' = \alpha'e^{2\omega}.}

\noindent That is, the only effect of switching from the string frame as far as the energy momentum tensor of a string gas is concerned is this field dependent redefinition of $\alpha'$  (c.f. \cite{sp2}). Notice that in the above, we have only derived the energy momentum tensor for a gas of strings with a fixed set of quantum numbers, indexed by \{$ \vec n, \vec w, N, p^2_d $\}, with a number density that will depend on these quantum numbers: $\mu = \mu_{\vec n, \vec w, N, p^2_d}$. In general our string gas will contain a whole host of different quantum numbers, and these will all have to be summed over to arrive at the correct energy momentum tensor. However it was shown in \cite{sp2} that if at some initial time we were in a situation of thermal equilibrium (defined through (\ref{nd})), then all the number densities, instead of being arbitrary are given by:

\eq{dist}{\mu_{\vec n, \vec w, N, p^2_d} = e^{-\beta \epsilon_{\vec n, \vec w, N, p^2_d}}\mu_{ref}e^{\beta\epsilon_{ref}},} 

\noindent where the subscript `ref' refers to some reference state, which we take to have zero energy, and $\epsilon$ is given by (\ref{tenergy})-- recalling that we have now dropped the tildes to indicate Einstein frame quantities. After careful consideration of what constitutes the thermal bath that couples these modes to each other, it was then shown in \cite{sp2} that by defining the following quantity (where we now generalize the result to include the contribution of the dilaton):

\begin{equation}
\label{z}
Z(\beta,a_i,b_a,\omega) := \mu_{ref} \sum_{\vec n,\vec w,N,p^2} e^{-\beta \epsilon(a_i,b_a,\omega)_{\vec n,\vec w,N,p^2}},
\end{equation} 

\noindent one could derive the components of the energy momentum tensor as follows:

\begin{eqnarray}
\label{p1i}
P_i &=& \frac{1}{\beta}a_i\frac{\partial Z}{\partial a_i} ~~=~~ \frac{1}{\beta}\frac{\partial Z}{\partial \lambda_i} ~~;~~ a_i = e^{\lambda_i},\\
\label{p1a}
P_a &=& \frac{1}{\beta}b_a\frac{\partial Z}{\partial b_a} ~~=~~ \frac{1}{\beta}\frac{\partial Z}{\partial \lambda_a} ~~;~~ b_a = e^{\lambda_a},\\
\label{r10}
E &=& \frac{-1}{\beta}{}\frac{\partial Z}{\partial \beta}.
\end{eqnarray}

\noindent where $P_\mu$ (and in a similar way, $E$) is defined through the equation:

\begin{equation}
\frac{P_\mu}{\sqrt{G}} = p_\mu, 
\end{equation}

\noindent with $p_\mu$ being the pressure along the $\mu^{th}$ direction. Notice that in this setup, our interaction lagrangian (\ref{int}) is given by:

\eq{mint}{S_{int} = -\int d^{D+1}x \sqrt{-g_{00}} E[g,\Phi],}

\noindent which corresponds to the framework of \cite{TV}. In \cite{sp1} and \cite{sp2}, it was argued that since the energy of a closed string in a given background is given by:

\eq{ecs}{\epsilon = \frac{\hat{\epsilon}}{\sqrt{\alpha'}},}

\noindent where $\hat\epsilon$ is a term of order unity for the lowest lying string states, and that since for all string theories, the Hagedorn temperature \cite{hag} is of order $\sqrt{\alpha'}$ \cite{mb}, that the summation in (\ref{z}) projects out all but the massless string states if we are even an order of magnitude below the Hagedorn temperature \cite{sp2}. Since our focus is on late times, we now have an operational meaning of such a time-- when the temperature of the universe has cooled to below an order of magnitude below the Hagedorn temperature. This is still considerably higher than the GUT scale, and so 'late times' in our framework refers only to the string scale. However this is only of passing interest to us, as we have argued that we shall only be interested in massless string modes from the outset. What this does indicate is that one can obtain these modes, demanded by phenomenological and theoretical consistency in a rather natural manner.  

\par
 
Turning now to the dilaton, one finds that the associated energy momentum tensor of the dilaton field is given by:

\begin{eqnarray}
\label{rd}
\rho_D &=& \frac{4}{(D-1)}\frac{1}{8\pi G_D} \frac{\dot{\Phi}^2}{2},\\
\label{pd}
p_D &=& \frac{4}{(D-1)}\frac{1}{8\pi G_D} \frac{\dot{\Phi}^2}{2},
\end{eqnarray}

\noindent where the pressure is the same for all directions (be they compact or not). Consider now the Einstein equation for the $\mu^{th}$ diagonal element of the metric tensor \cite{sp2}:

\eq{ee}{\ddot{a}_\mu + \dot{a}_\mu\Bigl(\sum_{\nu\neq \mu} \frac{\dot{a}_\nu}{{a_\nu}} \Bigr) = 8\pi G_Da_\mu\Bigr[p_\mu-\frac{1}{D-1}\sum_{\nu=1}^{D} p_\nu +\frac{1}{D-1}\rho \Bigl],}

\noindent where these equations are to be supplemented by the $00$ Einstein equation: 

\eq{00}{8\pi G_D\rho = \frac{1}{2}\sum_{\mu=1}^{D}\sum_{\nu \neq \mu}^{D}H_\mu H_\nu,}

\noindent where $H_\mu$ is the Hubble factor associated with the $\mu^{th}$ scale factor: $H_\mu = \dot{a}_\mu/a_\mu$. Notice that this equation is the only place where the dilaton field factors, as its contribution to the energy momentum tensor drops out of (\ref{ee}), as $\rho_D = p_D$. Inserting (\ref{rho})-(\ref{pa}) into (\ref{ee}), we find for the equations of motion for the scale factors for the compact directions as:

\eq{dc}{\ddot{b}_a + \dot{b}_a(\sum_j \frac{\dot{a_j}}{a_j} + \sum_{b \neq a}\frac{\dot{b}_c}{b_c}) = \sum_{m^2 = 0} \frac{8\pi G_D \mu_{0, \vec n, \vec w}b_a}{\sqrt{g_s}\epsilon_{\vec n, \vec w}}\Bigl[ \frac{n^2_a}{{b}^2_a} - \frac{w_a^2b_a^2}{\alpha'^2}e^{-4\omega} +  \frac{2}{(D-1)}[(w,\gamma w)\frac{e^{-4\omega}}{\alpha'^2} + \frac{e^{-2\omega}}{\alpha'}\{(n,w) + 2(N-1)\}] \Bigr].}
 
\noindent The Euler-Lagrange equation of motion for $\Phi$ is given by (recall that $\omega = 2\frac{(\Phi_0 - \Phi)}{(D-1)}$):

\eq{eld}{\ddot{\omega} + (\sum_j \frac{\dot{a_j}}{a_j} + \sum_{b}\frac{\dot{b}_c}{b_c})\dot{\omega} = \sum_{m^2 = 0} \frac{2}{(D-1)}\frac{8\pi G_D \mu_{0, \vec n, \vec w}}{\sqrt{g_s}\epsilon_{\vec n, \vec w}}\Bigl[ \frac{e^{-4\omega}}{\alpha'^2}(w,\gamma w) + \frac{e^{-2\omega}}{\alpha'}[(n,w) + 2(N-1)]\Bigr].}

\noindent The summation above is over all the massless quantum numbers which appear at the the given stage of the evolution of our background (recalling that (\ref{tenergy}) depends on the metric tensor), as previously discussed. Now we choose to begin at or around the self dual radius ($b_a \sim \sqrt{\alpha'}$), with $\Phi$ beginning at or around $\Phi_0$ (i.e. $\omega \sim 0$ as in (\ref{fix})). We compute the summation over modes which are massless at, and near this point in moduli space in appendix B, where we find this to yield the following equations of motion:

\eq{dc1}{
\ddot{b}_a + \dot{b}_a(\sum_j \frac{\dot{a_j}}{a_j} + \sum_{b \neq a}\frac{\dot{b}_c}{b_c}) = \frac{8\pi G_D \mu b_a}{\sqrt{g_s}|p_d|}(8p-4)\Bigl[ \frac{1}{{b}^2_a} - \frac{b_a^2}{\alpha'^2}e^{-4\omega} +  \frac{2}{(D-1)}[\sum_{c=1}^{p} b_c^2 \frac{e^{-4\omega}}{\alpha'^2} -p\frac{e^{-2\omega}}{\alpha'}] \Bigr],}
 
\eq{eld2}{\ddot{\omega} + (\sum_j \frac{\dot{a_j}}{a_j} + \sum_{c}\frac{\dot{b}_c}{b_c})\dot{\omega} = \frac{2}{(D-1)} \frac{8\pi G_D \mu}{\sqrt{g_s}|p_d|}(8p-4)\Bigl[ \frac{e^{-4\omega}}{\alpha'^2}\sum_{c=1}^p b_c^2 -p\frac{e^{-2\omega}}{\alpha'}\Bigr],}

\noindent where implicit in the above is that all the moduli are close to the values just discussed. Now we expand the above around:

\begin{eqnarray}
\label{bexp}
b_a &=& \sqrt{\alpha'}(1+\Gamma_a)~;~\Gamma_a \ll 1,\\
\label{phiexp}
\omega &;& \omega \ll 1.  
\end{eqnarray}

\noindent The equations of motion which result are:

\eq{gam}{\ddot{\Gamma}_a + dH\dot{\Gamma}_a + \frac{8\pi G_D \mu}{a^3\alpha'\sqrt{\alpha'}^{p}|p_d|}4(8p-4)\Bigl[(\Gamma_a - \omega) - \frac{1}{(D-1)}[\sum_{c=1}^{p}(\Gamma_c - \omega)]\Bigr] = 0,}

\eq{om}{\ddot{\omega} + dH\dot{\omega} + \frac{8\pi G_D \mu}{a^3\alpha'\sqrt{\alpha'}^{p}|p_d|}4(8p-4)\frac{1}{(D-1)}\Bigl[\sum_{c=1}^{p}(\omega - \Gamma_c)\Bigr] = 0.}

\noindent We now consider three separate cases of interest to us, namely when we take the dilaton to be fixed and the radii dynamical, when the radii are fixed, and the dilaton is dynamical, and when the radii and the dilaton are both taken to be dynamical. We will motivate the fact that we investiagate these cases seperately in what follows.

\subsection{Fixed Dilaton}

This case was already investigated in \cite{sp2}, and so our treatment here will be brief. One might imagine that the dilaton is stabilized by some mechanism external to string gas cosmology, either through potentials generated by higher order corrections, or some other mechanism such as supersymmetry breaking. Furthermore, one assumes this mechanism renders the dilaton massive, with a mass that is far greater than the mass of the radion (given by the coefficient of $\Gamma_a$ in (\ref{gam})). This allows us to ignore the fluctuations of the dilaton in (\ref{gam}), which results in the following equation of motion:

\eq{gamo}{\ddot{\Gamma}_a + dH\dot{\Gamma}_a + \frac{8\pi G_D \mu}{a^3\alpha'\sqrt{\alpha'}^{p}|p_d|}4(8p-4)\Bigl[\Gamma_a - \frac{1}{(D-1)}[\sum_{c=1}^{p}\Gamma_c ]\Bigr] = 0.} 

\noindent Clearly, if we study the fluctuation of a single mode (by setting $\Gamma_b = 0$ for all indices $b$ not equal to the mode we're interested in, then the above equation describes damped oscillations about the self dual radius ($\Gamma_a = 0$), with a mass squared given by the `spring constant':

\eq{sc}{m^2 = \frac{8\pi G_D \mu}{a^3\alpha'\sqrt{\alpha'}^{p}|p_d|}4(8p-4)\frac{(D-2)}{(D-1)}.}  

\noindent However to undertake a full stability analysis, one needs to compute the eigenvalues of the Hessian matrix corresponding to this coupled radion system, and show that they are all positive. Indeed, one can write the above driving term as $\partial_a V$, where upon defining $H = \partial_a\partial_b V$, we find that (up to an overall positive factor): 

\begin{equation}
\label{hess1}
H = \lambda \begin{pmatrix} D-2&-1&-1&\ldots&-1\\
-1&D-2&-1&\ldots&-1\\
-1&-1&D-2&\ldots&-1\\
\vdots&\vdots&\vdots&\ddots&\vdots\\
-1&-1&-1&\ldots&D-2
\end{pmatrix}
\end{equation}

\noindent where $\lambda$ is given by:

\eq{opp}{\lambda = \frac{8\pi G_D \mu}{a^3\alpha'\sqrt{\alpha'}^{p}|p_d|}\frac{4(8p-4)}{(D-1)}.}

\noindent Now $H$ is a $p$ by $p$ dimensional matrix, which has as its eigenvalues (in units of $\lambda$): 

\eq{heval}{\{D-1,D-1,...,D-1,D-p-1\},}

\noindent That is, $D-1$ appears with a degeneracy of $p-1$, and $D-p-1 = d-1$ appears once. Clearly, these are all positive and hence describe a stable equilibrium. If we consider the driving term of (\ref{dc1}) in its own right, that is by not restricting to small perturbations around the self dual radius, one finds that for the volume modulus, the driving term correpsponds to the potential indicated in fig 1.

\begin{figure}
\epsfig{figure=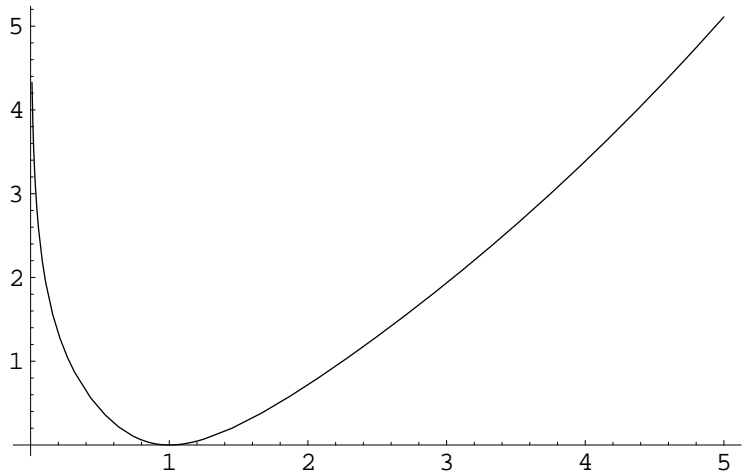}
\caption{Potential generated generated for the volume modulus ($V = (b/\alpha')^p$) in units of $\frac{8\pi G_d \mu_0}{\sqrt{g_s}}$ v.s. $b/\sqrt{\alpha'}$.}
\end{figure}

As found in \cite{sp2}, the non-compact dimensions undergo a period of FRW expansion after radion stabilization, with a radion mass given by the 'spring constant' in (\ref{gamo}), which upon using the relation:

\eq{newtred}{G_D = G_d(2\pi\sqrt{\alpha'})^p,}

\noindent and setting the scale of stringy physics to be at the Planck scale ($8\pi G_3 = 2\pi \alpha'$), yields a radion mass:

\eq{radmass}{m^2 \sim \frac{\mu_0}{a^3|p_d|},}

\noindent and a 4-d energy density:

\eq{raded}{\rho \sim \frac{\mu|p_d|}{a^3}.} 

\noindent We remind the reader that $|p_d|$ is the center of mass momentum (energy) of these massless strings. We have neglected factors of order unity in deriving the above. In \cite{sp2} it was shown that for all $|p_d| \leq 10^{-3}eV$ in the present epoch, one can evade fifth-force constraints whilst simultaneously ensuring that the energy density of this string gas (which looks like hot dark matter) remains below several orders of magnitude of the critical density. 

\subsection{Fixed Radii}

We now consider that somehow, the moduli corresponding to the extra dimensions has stabilized at the self-dual radius through some mechanism that doesn't couple to the dilaton (or at least contributes feebly to its dynamics), such as from the effects of some scalar field, or some other as yet unknown mechanism. The most obvious candidate mechanism, the effects higher dimensional branes in the presence of fluxes, turns out to couple to the dilaton, though it may be possible to arrange this coupling to not significantly affect the driving term (\ref{eld2}). The relevant point being that it is no less arbitrary to assume the stabilization of extra dimensions, than it is to assume the stabilization of the dilaton. Doing so, we find that the equation of motion for the dilation fluctuation becomes:

\eq{eld21}{\ddot{\omega} + (\sum_j \frac{\dot{a_j}}{a_j})\dot{\omega} = \frac{2}{(D-1)} \frac{8\pi G_D \mu_0}{\alpha'\sqrt{g_s}|p_d|}p(8p-4)\Bigl[ e^{-4\omega} -e^{-2\omega}\Bigr],}

\noindent which corresponds to motion in the potential (illustrated in fig 2.):

\eq{dpt}{V(\omega) = \frac{2}{(D-1)} \frac{8\pi G_D \mu_0}{\alpha'\sqrt{g_s}|p_d|}p(8p-4)\Bigl[ \frac{e^{-4\omega}}{4} -\frac{e^{-2\omega}}{2}\Bigr].}
   
\begin{figure}
\epsfig{figure=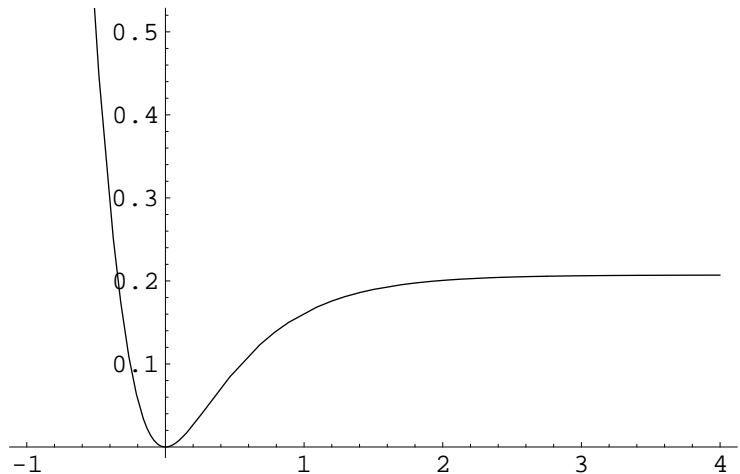}
\caption{Potential generated generated for $\omega$ in units of $\frac{16\pi G_d \mu_0p(8p-4)}{(D-1)\alpha'\sqrt{g_s}|p_d|}$ v.s. $\omega$.}
\end{figure}

\noindent The mass generated for the dilaton by the string gas is given by the curvature of this potential at its minimum, which can be read off (\ref{om}):

\eq{dilatonmass}{m^2_\omega = \frac{8\pi G_D \mu}{a^3\alpha'\sqrt{\alpha'}^{p}|p_d|}4p(8p-4)\frac{1}{(D-1)},}

\noindent again, using (\ref{newtred}) and setting the scale for stringy physics to be of the Planck scale ($2\pi\alpha' = G_3$), we find that:

\eq{dilmassf}{m^2_\omega \sim \frac{\mu_0}{a^3|p_d|},}

\noindent whereas the 4-d energy density of the string gas that generates this potential is given (as before) by (\ref{raded}). Note that all the features advertised in the introduction have made their appearance here. There was no need to worry about exactly where the miniumum of the potentials for the various moduli occur, as there is no need to fine tune when the form of the action for a string gas takes the form (\ref{mint}). We also see that the mass of the various moduli can be made large enough (to avoid fifth force type interactions) without overclosing the universe (see \cite{sp2} for a more detailed treatment). In taking stock of what we have accomplished, we see that we have stabilized the dilaton around its expectation value ($\omega = 0 \to \Phi = \Phi_0$), but we have no way of predicting precisely what this value is. Recall that the action (\ref{da}) undergoes an overall rescaling under an additive redefinition of the dilaton, where the observed Newtons constant is then given by:

\eq{obsnc}{8\pi G_d = \kappa^2_0 e^{2\Phi_0}.}

\noindent we find that whatever this $\Phi_0$ is, the dynamics of a gas of massless strings will stabilize the dilaton around this value, and hence render it constant. It does not have the ability to predict what the expectation value of the dilaton will be (a situation that as we will see, will change when we consider the full dynamics of the system). Certainly this seems to indicate something peculiar about the dynamics of our moduli stabilization framework, and as we will see shortly, is a first indication of the presence of a flat direction in moduli space. However at present, we can view this result as something of a boone, in that such a mechanism is capable of stabililzing the dilaton no-matter which particular corner of moduli space one finds themselves in. Of course one would like to be able to predict what the constants of nature are from a given moduli stabilization mechanism, and as such this is somewhat of an unsatisfactory situation. It is only of passing interest to us however, as the main focus of this report will be the case of when we consider the coupled dynamics of the radion-dilaton system, to which we turn our attention to now.        

\section{Coupled Dilaton-Radion Dynamics}

Recall the equations of motion for the fluctuations of the radion and the dilaton fields (\ref{gam})(\ref{om}):

\eq{gam00}{\ddot{\Gamma}_a + dH\dot{\Gamma}_a + \frac{8\pi G_D \mu}{a^d\alpha'\sqrt{\alpha'}^{p}|p_d|}4(8p-4)\Bigl[(\Gamma_a - \omega) - \frac{1}{(D-1)}[\sum_{c=1}^{p}(\Gamma_c - \omega)]\Bigr] = 0,}

\eq{om00}{\ddot{\omega} + dH\dot{\omega} + \frac{8\pi G_D \mu}{a^d\alpha'\sqrt{\alpha'}^{p}|p_d|}4(8p-4)\frac{1}{(D-1)}\Bigl[\sum_{c=1}^{p}(\omega - \Gamma_c)\Bigr] = 0.}

\noindent We have already see how the radion fields and the dilaton are separately stable. It is a straightforward generalization to study the case of when both sets of moduli are dynamical. To find out if the coupled system is stable, we have to find out if we are at a point in moduli space which corresponds to a global minimum. It seems straight forward to generalize the treatment above, where it seems that all we have to do is to compute the eigenvalues of the Hessian matrix. However there is a subtle issue here-- the driving term for the dilaton is given by:

\eq{ddrive}{\frac{\partial Z}{\partial \omega},}

\noindent whereas the driving term for the radii is given by (c.f. (\ref{z})-(\ref{mint})):

\begin{equation}
\label{diff}
\Bigl[a_\mu \frac{\partial }{\partial a_\mu} - \frac{1}{D-1}\Bigl(\sum_\nu a_\nu\frac{\partial }{\partial a_\nu} + \frac{\partial }{\partial \beta} \Bigr)\Bigr]Z.
\end{equation}

\noindent If one were to write the driving term for the radii as $\frac{\partial V}{\partial \Gamma}$  for some $V$, then in general this will not be the same $V$ that one derives the driving term for $\omega$ because of their different origins (wheras one comes from linear combinations of the Einstein equations, the other comes from the Euler Lagrange equations). However, were one to make the trivial rescaling:

\eq{rescale}{\Gamma_a = \sqrt{d-1}\gamma_a,}

\noindent where $d$ is the number of spatial dimensions, one can write the above driving terms as the partial derivative of a 'master potential'. The neccesity in doing this comes from the fact that were we to naively take the second derivatives of the driving term above, one finds that $\partial_\omega\partial_{\Gamma_a} V \neq \partial_{\Gamma_a}\partial_\omega V$. This rescaling allows us to write the above driving terms as the derivative of the appropriate potential. From this, we find that, up to an overall positive factor:

\begin{eqnarray}
\partial_a\partial_\omega V = \partial_\omega\partial_a V&=& -\frac{\sqrt{d-1}}{(D-1)},\\
\partial_\omega^2 V &=& \frac{p}{(D-1)},\\
\partial_a\partial_b V &=& -\frac{1}{(D-1)},\\
\partial_a^2 V &=& \frac{(D-2)}{(D-1)},
\end{eqnarray}

\noindent which corresponds to the $p+1$ by $p+1$ dimensional Hessian matrix (again up to the positive factor (\ref{opp})):

\begin{equation}
\label{hess}
H_F = \lambda
\begin{pmatrix}
D-2&-1&-1&\ldots&-1&-{\sqrt{d-1}}\\
-1&D-2&-1&\ldots&-1&-{\sqrt{d-1}}\\
-1&-1&D-2&\ldots&-1&-{\sqrt{d-1}}\\
\vdots&\vdots&\vdots&\ddots&\vdots\\
-1&-1&-1&\ldots&D-2&-{\sqrt{d-1}}\\
-{\sqrt{d-1}}&-{\sqrt{d-1}}&-{\sqrt{d-1}}&\ldots&-{\sqrt{d-1}}&p
\end{pmatrix}.
\end{equation}
  
\noindent This matrix has eigenvalues $\lambda(D-1)$, which appears with degeneracy of $p$, and 0, which appears once (in fact the corresponding eigenvalue for such a p+1-dimensional matrix is $\lambda(-p-d+D)=0$). The appearance of a zero mode to this system implies that one degree of freedom in this coupled system is not stabilized (albeit marginally). In fact, one can directly obtain the normal modes of the system from (\ref{gam00})(\ref{om00}):

\eq{norml}{\ddot{\Lambda}_a + dH\dot{\Lambda}_a + \frac{8\pi G_D \mu}{a^d\alpha'\sqrt{\alpha'}^{p}|p_d|}4(8p-4)\Lambda_a = 0~~;~~ \Lambda_a := \Gamma_a - \omega,}

\eq{normk}{\ddot{\kappa} + dH\dot{\kappa} = 0~~;~~ \kappa = \sum_{a=1}^p \Gamma_a + (d-1)\omega,}

\noindent from which we conclude that the individual radial moduli ($\Gamma_a$) tend towards the dilaton (i.e. $\Gamma_a - \omega \to 0$) quite robustly, which, from the definition of $\kappa$, results in $\kappa$ tending towards $(D-1)\omega$, which now no longer has a mass. This is not to say that this modulus is free to run away, as according to (\ref{normk}), one can solve for $\kappa$ to find that:

\eq{kappaeom}{\kappa(t) = c\int_{t_0}^{t}\frac{1}{a^d(t)} dt,}  
 
\noindent which implies that in a universe with expanding non-compact dimensions (which is easily obtained in our framework \cite{sp1}\cite{sp2}), this modulus terminally approaches some finite value. However, this is far from satisfying to say the least, as this modulus is no longer robustly held at any particular value and fluctuations and interactions are likely to shift the value of this modulus significantly through time.

\par

One can quite easily understand the physical origins of this zero mode. Recalling the energy of a closed string (\ref{tenergy}) in the given background:

\begin{equation}
\nonumber
\epsilon = \sqrt{|{p}_{d}|^2 + (n,\gamma^{-1} n) + \frac{e^{-4\omega}}{\alpha'^2}(w,\gamma w) + \frac{e^{-2\omega}}{\alpha'}[2(n,w) + 4(N-1)],}
\end{equation}

\noindent we see that if instead of stablizing at $0$, were $\omega$ to stabilize at some other value, say $\omega_0$, then modes which were massless for $\omega = 0$ at the self dual radius are now massless at $\omega = \omega_0$ at the radius defined by $b_a = \sqrt{\alpha'}e^{\omega_0}$. Hence were we to begin at this point in moduli space, and compute the sum over massless modes in our equations of motion (as in how we arrived at (\ref{gam00})(\ref{om00})), one would end up formally identical equations for the fluctations about this point in moduli space, and obtain the same results for our stabilizty analysis. This would certainly indicate that we have a flat direction in the dynamics of our coupled system. 

\par
 
Before we attempt to resolve this unsatisfactory state of affairs, we wish to point out that were the dilaton to receive a potential due to effects external to SGC (i.e. through higher order corrections), then one retrieves a stable equilibrium for our system {\it for any value of the stabilized dilaton $\omega = \omega_0$}, where the extra dimensions stabilize at $b_a = \sqrt{\alpha'}e^{\omega_0}$. That is, were some external potential to stabilize the dilaton, the Hessian matrix at the stabilized point in moduli space would then read:

\begin{equation}
\label{hess2}
H_F=\lambda\begin{pmatrix}
D-2&-1&-1&\ldots&-1&-{\sqrt{d-1}}\\
-1&D-2&-1&\ldots&-1&-{\sqrt{d-1}}\\
-1&-1&D-2&\ldots&-1&-{\sqrt{d-1}}\\
\vdots&\vdots&\vdots&\ddots&\vdots\\
-1&-1&-1&\ldots&D-2&-{\sqrt{d-1}}\\
-{\sqrt{d-1}}&-{\sqrt{d-1}}&-{\sqrt{d-1}}&\ldots&-{\sqrt{d-1}}&p + cV''(\omega)
\end{pmatrix},
\end{equation}
  
\noindent where the only difference is in the last diagonal entry. In the above, $c$ is some positive factor (the overall factor for the string gas contributions to the Hessian will not be the same as that for the dilaton stabilization mechanism (\ref{opp})). This matrix generically has all of its eigenvalues as positive for any positive $V''$. For instance, for 6 extra dimensions in $D=10$, one finds that the Hessian matrix has the eigenvalues (in units of $\lambda$):

\eq{hev}{\{9,9,9,9,9,\frac{1}{2}(9+cV'' - \sqrt{81 + 6cV'' + (cV'')^2}),\frac{1}{2}(9+cV'' + \sqrt{81 + 6cV'' + (cV'')^2}) \},}    

\noindent which are all positive. Hence we conclude that one can consistently stabilize extra dimensions in string gas cosmology independent of any particular (external) dilaton stabilization mechanism. This ameliorates the difficulties encountered in \cite{ab}, which studied precisely this aspect of the problem, but neglected the effects of massless string states. 

\par

Returning however to the issue at hand, which is to explore ways in which some mechanism within SGC might lift the zero mode of the coupled dilaton-radion system, we offer a quick recap of the possibilities raised previously. Firstly, one could introduce higher dimensional branes and turn on fluxes, in the hope that this can somehow enhance the stabilization of the extra dimensional moduli but not the dilaton (or vice versa), in which case we would end up with the two separate situations considered in the previous section. Alternatively, one could introduce branes and fluxes and not aim for preferential enhancement of the stabilization mechanisms of one of the two sets of moduli, but instead see if doing so can somehow otherwise lift the zero mode (as we have seen, obtaining even a small additional positive component to the last diagonal entry in (\ref{hess2}) suffices). However there are compelling reasons for us to focus solely on one particular possibility-- namely the introduction of a gas of D-strings.

\section{D-strings} 

In the introduction, we made a rather detailed case as to why massless string modes should be the only concern of string gas cosmology if it aspires to be relevant to the present epoch. Consider then what it would mean to introduce branes of various dimensions into our framework. Since we do not yet know how to quantize any extended object other than strings, we will have to content ourselves with a purely classical treatment of such objects, for which the lowest energy configurations have energies:

\eq{dbe}{E = T_p\times Vol}  

\noindent where $Vol$ is the worldvolume of such an object, and $T_p$ is the tension of such a p-dimensional object (typically the p'th power of the string tension). We see that introducing a gas of such objects would run counter to the spirit of what we have done so far, in that we would then be introducing objects which have a significant energy density, calling into question any applicability to late time cosmology, and possibly taking us away from the regime of the low energy effective theory (\ref{se}) in which we are working in. Hence if we are to be consistent with our own demands, we should ignore these objects altogether. However, we find that there is one class of D-branes which evades the objections just made, namely D-strings. Not only are D-strings quantizable (in certain backgrounds, with caveats), they also have the same massless excitations as the F-string \cite{pol}, and offer a window of opportunity to lift the zero mode uncovered in the previous section as they couple to the dilaton in a different manner than F-strings. This follows from the action for D-string in the absence of fluxes (in the string frame):

\eq{dstra}{S = -\frac{T_1}{2\pi}\int d^2\sigma e^{-\Phi}\sqrt{-h}}    

\noindent where $T_1/2\pi$ is related to the D-string tension ($:= \tau_1$) in a background with the dilaton assuming the value $\Phi_0$ through the relation:

\eq{dst}{\tau_1 = \frac{e^{-\Phi_0}T_1}{2\pi}}

\noindent We will elaborate on the tension of the D-string further on. In what follows, we will be using the fact that as mentioned, one can indeed quantize the D-string in certain backgrounds. As a solitonic object in string theory, whether one can do this or not is a highly non-trivial question. However it should be clear that for a constant dilatonic background, (\ref{dstra}) describes the same action as that for the F-string, with a corresponding redefinition of $\alpha'$. One should certainly be able to quantize this action and recover the spectrum of states (which follows from the constraint algebra alone), if one can do this for the F-string. It was shown in \cite{hk} that one can indeed covariantly quantize the superstring, provided the form fields satisfied certain conditions. In \cite{kami} the canonical equivalence of the constraint algebra of the D-string and the type IIB string is established, which implies that these two objects have the same spectrum. Since we have endeavoured to keep the background theory in which our framework is established arbitrary, we do not consider this issue further, taking as an assumption that one can indeed quantize D-strings in the corner of moduli space we are considering. We argue in the appendix that for situations in which the dilaton field changes much, much  slower than the string scale (defined by $\dot \Phi \ll 1/2\pi\alpha'$-- which we show later holds to an excellent approximation in our model), the only change in the D-string gas energy momentum tensor after switching to the Einstein frame (\ref{ct}), compared to that for the F-string gas (\ref{rho})-(\ref{redef}), is the replacement $\alpha' \to e^\Phi/T_1$: 
  
\begin{eqnarray}
\label{rhod}
\rho &=& \frac{\mu_D\epsilon}{\sqrt{{g}_s}},\\
\label{pid}
p^i &=& \frac{\mu_D}{\sqrt{{g}_s}\epsilon}|p_{d}|^2/d,\\
\label{pad}
p^a &=& \frac{\mu_D}{\sqrt{{g}_s}\epsilon}[\frac{n_a^2}{{b}_a^2} - \frac{w_a^2 {b}_a^2e^{-4\omega}}{(e^\Phi/T_1)^2}],
\end{eqnarray}

\eq{tdenergy}{\epsilon = \sqrt{|{p}_{d}|^2 + (n,\gamma^{-1} n) + \frac{e^{-4\omega}}{(e^\Phi/T_1)^2}(w,\gamma w) + \frac{e^{-2\omega}}{(e^\Phi/T_1)}[2(n,w) + 4(N-1)]}.}

\noindent Further issues concerning the consistent introduction of a D-string gas are considered in section V. Recalling the relation (\ref{fix}), one can rewrite $\Phi$ as:

\eq{redp}{\Phi = \Phi_0 - \frac{D-1}{2}\omega,}

\noindent and by defining the following quantity:

\eq{alalt}{\hat\alpha:= e^{\Phi_0}/T_1,}

\noindent we can rewrite the above expression for the energy of a single string as:

\eq{ealt}{\epsilon = \sqrt{|{p}_{d}|^2 + (n,\gamma^{-1} n) + \frac{e^{(D-5)\omega}}{\hat\alpha^2}(w,\gamma w) + \frac{e^{(D-5)\omega/2}}{\hat\alpha}[2(n,w) + 4(N-1)]}.}

\noindent where the numerologically peculiar factor $D-5$ will turn out to have little consequence. A gas of D-strings (with a fixed set of quantum numbers) will yield an energy momentum tensor (in the Einstein frame) that can be derived from the action:

\eq{Dact}{S= -\int d^{D+1}x \sqrt{g_{00}}\mu_{D,\vec n,\vec w ,N}\sqrt{|{p}_{d}|^2 + (n,\gamma^{-1} n) + \frac{e^{(D-5)\omega}}{\hat\alpha^2}(w,\gamma w) + \frac{e^{(D-5)\omega/2}}{\hat\alpha}[2(n,w) + 4(N-1)]},}

\noindent where the subscript $D$ on $\mu$ indicates that this number density is specific to the D-string gas, which for massless states we take as equal to the number density for the F-string gas. Just as before, if we realise that one has contributions from each massless state, then one has to perform a sum over massless modes to arrive at the correct energy momentum tensor (see discussion immediately following (\ref{dist})). The equations of motion that result are as follows: 

\eq{Ddc}{\ddot{b}_a + \dot{b}_a(\sum_{\mu \neq a} H_\mu) = \sum_{m^2 = 0} \frac{8\pi G_D \mu_{D,\vec n, \vec w}b_a}{\sqrt{g_s}\epsilon_{\vec n, \vec w}}\Bigl[ \frac{n^2_a}{{b}^2_a} - \frac{w_a^2b_a^2}{\hat\alpha^2}e^{(D-5)\omega} + \frac{2}{(D-1)}[(w,\gamma w)\frac{e^{(D-5)\omega}}{\hat\alpha^2} + \frac{e^{(D-5)\omega/2}}{\hat\alpha}\{(n,w) + 2(N-1)\}] \Bigr],}

\eq{Deld}{\ddot{\omega} + (\sum_\mu H_\mu)\dot{\omega} = -\sum_{m^2 = 0} \frac{(D-5)}{2(D-1)}\frac{8\pi G_D \mu_{D, \vec n, \vec w}}{\sqrt{g_s}\epsilon_{\vec n, \vec w}}\Bigl[ \frac{e^{(D-5)\omega}}{\hat\alpha^2}(w,\gamma w) + \frac{e^{(D-5)\omega/2}}{\hat\alpha}[(n,w) + 2(N-1)]\Bigr].}

\noindent We now make the observation that at a generic point in moduli space, quantum numbers which render the F-string massless are not the same as those which render the D-string massless (a detailed treatment of what follows is given in appendix B). For instance, at the self-dual radius ($\gamma_{ab} = \alpha'\delta_{ab}$), with the dilaton stabilized around $\Phi_0$ (i.e. $\omega = 0$), the mass of the F-string is given by (c.f. (\ref{tenergy})):

\eq{mfs}{\alpha'm^2 = (n+w,n+w) + 4(N-1),} 

\noindent from which we catalogue the classes of states which are massless (recalling that the above formula must be supplemented with the level matching constraint $N +(n,w) \geq 0$): 

\begin{center}
\begin{tabular}{c|ccc|}
$N~$&$(n,n)~$&$(w,w)~$&$(n,w)~$\\
\hline
1&0&0&0\\
1&1&1&-1\\ 
0&1&1&1\\
0&1&3&0\\
0&3&1&0\\
0&2&2&0\\
0&4&0&0\\
0&0&4&0\\
\end{tabular}
\end{center}

\noindent Whereas the mass of the D-string at this point in moduli space is given by: 

\eq{mmds}{\alpha'm^2= (n,n) + \Bigl(\frac{\alpha'}{\hat\alpha}\Bigr)^2(w,w) + 2\Bigl(\frac{\alpha'}{\hat\alpha} \Bigr)[(n,w) + 2(N-1)],}

\noindent where $\hat\alpha$ is defined in (\ref{alalt}). We see from this that unless $\hat\alpha = \alpha'$, the quantum numbers tabulated above are not going to correspond to massless D-strings. Since the stabilization mechanism uncovered previously \cite{sp2} relied so heavily on the features of modes with these quantum numbers, right down to their degeneracies and behaviour away from the self-dual point (see appendix B), we take as a non-trivial assumption that we are in a region in moduli space where indeed $\hat\alpha = \alpha'$. Doing so allows us to sum over the same set of quantum numbers in (\ref{Ddc}) and (\ref{Deld}), and as we are about to see shortly, this will have some striking consequences. We will return to address the non-trivial point of whether or not one can consistently set $\hat\alpha$ to be equal to $\alpha'$ further on. For now, we see that the effect of introducing a gas of D-strings at this particular point in moduli space is to contribute to the equations of motion (\ref{Ddc})(\ref{Deld}) as:

\eq{Ddc1}{\ddot{b}_a + \dot{b}_a(\sum_\mu H_\mu) = \frac{8\pi G_D \mu_D b_a}{\sqrt{g_s}|p_d|}(8p-4)\Bigl[ \frac{1}{{b}^2_a} - \frac{b_a^2}{\alpha'^2}e^{(D-5)\omega} +  \frac{2}{(D-1)}[\sum_{c=1}^{p} b_c^2 \frac{e^{(D-5)\omega}}{\alpha'^2} -p\frac{e^{(D-5)\omega/2}}{\alpha'}] \Bigr],}
 
\eq{Deld2}{\ddot{\omega} + (\sum_\mu H_\mu)\dot{\omega} = -\frac{(D-5)}{2(D-1)} \frac{8\pi G_D \mu_D}{\sqrt{g_s}|p_d|}(8p-4)\Bigl[ \frac{e^{(D-5)\omega}}{\alpha'^2}\sum_{c=1}^p b_c^2 -p\frac{e^{(D-5)\omega/2}}{\alpha'}\Bigr].}

\noindent Of course, implicit in the above is that all the moduli are close to the values just discussed (all radii are close to the self-dual radius and $\Phi \approx \Phi_0$, consisitent with $\hat\alpha = \alpha'$). Just as before, we expand the above around:

\begin{eqnarray}
\label{Dbexp}
b_a &=& \sqrt{\alpha'}(1+\Gamma_a)~;~\Gamma_a \ll 1,\\
\label{Dphiexp}
\omega &;& \omega \ll 1.  
\end{eqnarray}

\noindent The resulting equations of motion are:

\eq{Dgam}{\ddot{\Gamma}_a + dH\dot{\Gamma}_a + \frac{8\pi G_D \mu_D}{a^3\alpha'\sqrt{\alpha'}^{p}|p_d|}4(8p-4)\Bigl[(\Gamma_a + \frac{(D-5)}{4} \omega) - \frac{1}{(D-1)}[\sum_{c=1}^{p}(\Gamma_c + \frac{(D-5)}{4}\omega)]\Bigr] = 0,}

\eq{Dom}{\ddot{\omega} + dH\dot{\omega} + \frac{8\pi G_D \mu_D}{a^3\alpha'\sqrt{\alpha'}^{p}|p_d|}4(8p-4)\frac{(D-5)}{4(D-1)}\Bigl[\sum_{c=1}^{p}(\frac{(D-5)}{4}\omega + \Gamma_c)\Bigr] = 0.}

\noindent By performing the same rescaling as in (\ref{rescale}), one can write the driving terms of this coupled system of equations as the derivatives of an appropriate potential, from which we arrive at the Hessian matrix (up to the same positive factor (\ref{opp}) that appears in front of (\ref{hess})):

\begin{equation}
\label{hess3}
H_D=
\lambda\begin{pmatrix}
D-2&-1&-1&\ldots&-1&{\frac{D-5}{4}\sqrt{d-1}}\\
-1&D-2&-1&\ldots&-1&{\frac{D-5}{4}\sqrt{d-1}}\\
-1&-1&D-2&\ldots&-1&{\frac{D-5}{4}\sqrt{d-1}}\\
\vdots&\vdots&\vdots&\ddots&\vdots\\
-1&-1&-1&\ldots&D-2&{\frac{D-5}{4}\sqrt{d-1}}\\
{\frac{D-5}{4}\sqrt{d-1}}&{\frac{D-5}{4}\sqrt{d-1}}&{\frac{D-5}{4}\sqrt{d-1}}&\ldots&{\frac{D-5}{4}\sqrt{d-1}}&p\frac{(D-5)^2}{16} 
\end{pmatrix}
\end{equation}

\noindent However, we have stipulated that both F-strings and D-strings will be present, so in addition to the driving terms generated by the D-string gas in the above, we will also have the driving terms generated by the F-string gas (\ref{gam}), (\ref{om}), which will result in the Hessian matrix for the combined system to be $H_{D+F} = H_F + H_D$, where $H_F$ is given by (\ref{hess}). It is here that we find the result we have been looking for: the eigenvalues of $H_{D+F}$ are all positive, inspite of both $H_F$ and $H_D$ individually possesing a zero mode. Hence all moduli fields have a stable equilibrium at the point in moduli space we have identified. For the specific case of $D=9$, $p=6$, we find that the eigenvalues of the Hessian matrices we consider are:

\begin{eqnarray}
\label{hf}
H_F &:& \{8, 8, 8, 8, 8, 8, 0\}\\
H_D &:& \{8, 8, 8, 8, 8, 8, 0\}\\
H_{F+D} &:& \{16, 16, 16, 16, 16, 12, 4\},
\end{eqnarray} 

\noindent where all eigenvalues are in units of the overall positive factor (\ref{opp}). We wish to point out that the fact that $H_F$ and $H_D$ have identical spectra is only a numerological coincindence for $D=9,p=6$. One can investigate the stability of the coupled D-string and F-string system in any number of dimensions (albeit numerically), and find that we always lift the zero mode we uncovered previously. For instance, for $D=25$, $p=22$, we find:

\begin{eqnarray}
\label{hf25}
H_F &:& \{24,24,24,...,24,24,0\}\\
H_D &:& \{552,24,24,...,24,24,0\}\\
H_{F+D} &:& \{ 12(24+\sqrt{565}),48,48,...,48,48,12(24-\sqrt{565})\}
\end{eqnarray}

\noindent Where the last eigenvalue in $H_{F+D}$ evaluates at $2.76\lambda$. What this implies is that all of the moduli we consider, are stabilized at the self-dual radius ($b_a = \sqrt{\alpha'}$), with the dilaton stabilized at the point determined by the condition $\hat\alpha = \alpha'$. Before we consider the phenomenology of this conclusion, we should return to address the issue of whether or not one can consistently set $\alpha' = \hat\alpha$. 
\par
We know in general that the D-string tension is related to the F-string tension ($= 1/2\pi\alpha'$) through the following relation,  which is exact and is a consequence of supersymmetry \cite{pol}:

\eq{trel}{\tau_1 = \frac{e^{-\Phi_0}}{2\pi\alpha'}.}

\noindent Therefore, one can only set the two tensions to be equal to each other at the self S-dual point ($e^{\Phi_0}=1$), from which we conclude that at weak coupling (which is the only regime in which the low energy effective theory (\ref{dact}) is valid), it is impossible for the two tensions to be equal to each other. Hence the stabilization mechanism we just proposed would seem to be untenable. In fact, for a generic point in moduli space corresponding to weak coupling, we have the relation:

\eq{wc}{\frac{\alpha'}{\hat\alpha} \gg 1.}

\noindent Thus one will never have a situation in which the same quantum numbers describe massless excitations of both the wound D-string and the wound F-string. It is interesting however to determine what the effects of the massless states of the D-string are, at a generic point in moduli space consistent with the weak copuling limit. Clearly states with the quantum numbers $N=1, \vec n, \vec w = 0$ correspond to massless D-strings as they do to massless F-strings, and correspond to unwound gravitons, which do not directly contribute to the dynamics of the radion fields or the dialton (c.f. (\ref{Ddc}),(\ref{Deld}) for the D-string and (\ref{dc}), (\ref{eld}) for the F-string). If were we to start out at the self-dual radius at weak coupling, then in general the only D-string states which are massless correspond to unwound graviton states $N=1,\vec n,\vec w=0$. However, for the specific case when $\alpha'/ \hat\alpha = q;~ q ~\epsilon ~\mathbb Z \gg 1$, it can be shown that the only massless states are those with quantum numbers:

\eq{wcdms}{(n,n) = 4q~;~N=0, \vec w = 0,} 
 
\noindent the specific realisations (and degeneracies) of which, depend on the number of extra dimensions we are considering. We see that quite generally, these states consist of only momentum modes and from (\ref{Ddc}) and (\ref{Deld}), we see that they have the net effect of causing the radion fields and the dilaton to run away. The run away behaviour for the dilaton field is such that it drives the dilaton towards stronger coupling. In particular, if we start out at weak coupling, the dynamics of the system tend to drive it towards the self dual coupling point-- where one would naively expect to find moduli stabilization from the results just uncovered. Quite obviously however, one expects qualitatively new physics to appear at the self S-dual point in moduli space, and so our framework is expected to break down at this point as it is highly unlikely that these new effects will conspire in such a way that it leaves our framework alone. 

\par

Since the issue of the D-string tension (\ref{trel}) is not up for debate, it might seem as if we have failed in our efforts to find a convincing moduli stabilization mechanism withing the framework of SGC. However, we have to keep in perspective that not only are we attempting to solve the moduli problem, we are attempting to solve it in a way that is relevant to the present epoch. Specifically, we are proposing a model which to our knowledge is the only attempt at stringy cosmology which makes contact with, and is operational in the present stage in our universe's evolution. We made a case for why one should only consider massless string degrees of freedom in the introduction with this in mind, and with this in mind presently, one should certainly expect us to be in a situation where we are in a regime with broken supersymmetry. It is almost immediately built into our framework that we are in a situation of broken supersymmetry, as introducing a gas of massless D-strings, whose constituents are moving around at the speed of light with respect to each other, are not going to be BPS (many bodied) states. There are exactly two consequences of this observation on what we have done so far. Firstly, since we have restricted ourselves to the bosonic sector of whatever string theory one might wish to apply this model to, the only difference for our framework will be in the critical dimension in which we wrote down the low energy effective action (\ref{dact}). Since we have deliberately left this dimensionality arbitrary, we do not have to change any aspect of our setup. However, one can make the argument that one can still consider spacetimes with the usual dimensionality (i.e. either 10 or 11) in spite of broken supersymmetry, for if we can indeed show stabilization at the string scale that is consistent with the low energy framework, any external physical probe of the scale of these extra dimensions will neccesarily have to be of a very high energy. At such energies, one certainly expects supersymmetry to be restored. This allows us to conclude that at most, the critical number of dimensions dictated by superstring theory are observable, inspite of the present epoch having broken supersymmetry. However this last point is in no way essential to our argument, as we have shown moduli stabilization in any number of spatial dimensions (D=10, 11, 26...), provided one can find a point in moduli space where the D-string and the F-string tension are equal to each other. 

\par

The second consequence of broken supersymmetry is by far the most important to us, for in this case, the relation (\ref{trel}) is longer true. In its place, is the relation (derived in D=25, by comparing F-string exchange amplitudes between two D-strings) \cite{pol}\cite{cj}:

\eq{bst}{\tau_1 = \frac{\sqrt\pi}{16\kappa_{0,D}e^{\Phi_0}}(4\pi^2\alpha')^5,}

\noindent where $\kappa_{0,D}$ is the normalization of the D-dimensional low energy effective action (\ref{dact}), which is determined by studying dualities between various string theories, and is proportional to $\alpha'$ to the appropriate power. The observed D-dimensional Newton's constant is given by:

\eq{newtd}{8\pi G_D = \kappa_{0,D}^2e^{2\Phi_0}.}

\noindent If one were to have compactified (or achieved radial moduli stabilization of) the p extra dimensions, $\kappa_0$ satisfies the relation:

\eq{kap}{\kappa_{0,D}^2 = \kappa_{0,d}^2\alpha'^{p/2} Vol T_p,}
 
\noindent where $T_p$ is the volume of the p compactified dimensions in units of $\alpha'^{p/2}$ ($T_p = (2\pi)^p$ for toroidal compactifications), and the observed d-dimensional Newtons constant (d being the number of non-compact dimensions, which we will soon take to be 3) is given by:

\eq{ddnc}{8\pi G_d = \kappa_{0,d}^2e^{2\Phi_0}.}

\noindent We immediately see from substituting (\ref{kap}) into (\ref{bst}), that the D-string tension depends on the compactification mechanism we have through the new relation:

\eq{bst2}{\tau_1 = \frac{\sqrt\pi(4\pi^2\alpha')^5}{16\kappa_{0,d}\alpha'^{p/4} (Vol T_p)^{1/2} e^{\Phi_0}},} 

\noindent which indicates the possibility of tuning the tension of the D-string, which depends on the string coupling, the compactification scheme and the particular string theory we are working in. Returning to the original expression (\ref{bst}), let us now assume that we start out at a value for $\Phi_0$, such that the D-string tension is equal to the F-string tension:

\eq{tune}{\tau_1 = \frac{1}{2\pi\alpha'} ~\to~ e^{\Phi_0} = \frac{\sqrt\pi 2\pi\alpha'}{16\kappa_{0,D}}(4\pi^2\alpha')^5.}

\noindent The overall coupling $\kappa_{0,D}$ that appears in the low energy effective action (\ref{dact}) is given for type I, IIA, IIB and Heterotic string theories (in $D=9$) as \cite{pol}:

\eq{norm}{\kappa_{0,9}^2 = \frac{1}{2}(2\pi)^7\alpha'^4.}

\noindent We thus conclude from the general relationship (\ref{kap}), that when we are in a situation where all extra dimensions are toroidally compactified at the self dual radius, that the relation (\ref{bst}) allows us to abstract from (\ref{tune}) the value of the string coupling where the D-string and the F-string tension are equal to each other:

\eq{scp}{e^{\Phi_0} = \frac{1}{16}.} 

\noindent In this case, one would have the situation where $\alpha' = \hat\alpha$ consistent with weak coupling, and massless excitations of the D-string would have the same quantum numbers at the self-dual radius as those of the F-string. We conclude that in this case, it is certainly possible for a gas of massless strings to stabilize all the moduli we have considered. In fact, as argued earlier, if we were to begin away from this point in moduli space, the dynamics of the coupled dilaton gravity-string gas system would drive the moduli fields to this point, where stabilization would result. We also wish to take note of the fact that the usual instability of D-strings in bosonic string theory is no longer at issue when the tensions of the two objects are equal, as the distinction between which one is fundamental becomes blurred, as there are no longer any obvious decay channels for the D-string. Note that not only does this mechanism make a prediction for the string coupling constant:

\eq{gs}{g_s = \frac{1}{16},}    

\noindent it also predicts the observed 3-dimensional Newtons constant (through the appropriate dimensional reduction of (\ref{norm})) in terms of the string scale:

\eq{rdnc}{8\pi G_3 = \frac{2\pi\alpha'}{32},}

\noindent which is certainly in the right ball park if we set the scale of stringy physics to be at or around the Planck scale. The phenomenology of this stabilization is exactly the same as the special cases we considered in the previous sections. All of the moduli fields, when expressed in terms of normal modes will have masses given by the curvature of the stabilizing potential at the minimum, given by (\ref{opp}), which upon using the relation (\ref{newtred}) gives us masses for the moduli as:

\eq{modmassa}{m^2 \sim \frac{\mu_0}{a^3|p_d|}.}  

\noindent Similarly, the contribution of the string gas to the energy momentum tensor will go like:

\eq{cosmomass}{\rho \sim \frac{\mu_0|p_d|}{a^3},}

\noindent from which it was shown in \cite{sp1} and \cite{sp2}, that for $|p_d| \leq 10^{-3}eV$, it is possible to evade fifth force constraints whilst simultaneously ensuring that the energy density of the string gas is substantially less than the critical density of the universe. Notice that the only parameter we have to tune this model is the center of mass energy of these strings ($|p_d|$), and even this only has to satisfy a rather mild bound than to be tuned in any sense.
Everything else in this framework determined by the requirement that we are in a regime where low energy effective theory is operative, and hence only massless modes will be contributing to the dynamics of spacetime. It is through the manifold remarkable properties of massless modes that we obtained our desired result. We wish to make the remark that string modes which have quantum numbers associated with massive modes not only have an inconsistent phenomenology, but  are also not likely to be around in any significant number in the present epoch, in addition to not leading to any stabilization mechanism. This is because any potential generated by massive string states refuses to admit terms which compete with one another such that a minimum results. It even appears that a no-go theorem to this effect, for moduli stabilization using massive modes is afoot \cite{tbab}. However, pressing on with our goal of stabilizing all the moduli of our given compactifications, we now turn our attention to stabilizing the shape moduli for the toroidal geometries we've been examining. Along the way we'll tie up a few loose ends, after which we'll conclude this report.  

\section{Shape Stabilization, Consistency 
Checks and Summary}

Since the issue of stabilizing the shape moduli was already considered in \cite{edna}, we do not consider the issue in great detail here as the mechanism through which it stabilizes is identical in our setup. In general, when we refer to the shape moduli of the toroidal compactification we are studying here, we mean the off-diagonal components of the metric for the internal manifold. Allowing the shape moduli to be dynamical means that our toroidal metric has components such that:

\eq{shape}{\gamma_{ab} \neq 0~,~\forall a\neq b.}

\noindent For illustrative purposes, we consider the simple example of a compactification on a two torus. How the results that follow generalize to higher dimensional tori will be indicated as we progress. In this case, the metric of the torus has the form:

\eq{tormet}{\gamma_{ab} = \begin{pmatrix} b_1& c\\c&b_2\end{pmatrix}.}

\noindent If we consider turning on the shape moduli such that they are close to vanishing, we find that to first order these moduli fields only make a difference to the off-diagonal elements of the Einstein Tensor (a result that is also valid for higher dimensional tori). Hence all of the results just derived for the other moduli fields are still true-- the shape moduli decouple completely from all other moduli to first order. In the case we are considering, we only have one extra Einstein equation to consider, namely the off-diagonal equation for the internal dimensions. This equation will read:

\eq{shapeq}{\ddot c + (dH -\mathcal H_1 - \mathcal H_2)\dot c + \frac{8\pi G_D}{2}\bigl[ c(\sum_\mu p_\mu - \rho) - 4T_{12}\bigr] = 0,}  

\noindent where we picked the indices $1,2$ for the compact directions. The term $(\sum_\mu p_\mu - \rho)$ vanishes around the moduli stabilization point, and deviations from this point result in terms which are linear in the infinitesimal fluctuations of the radial and dilaton moduli fields. Since $c$ itself is infinitesimal, this term is negligable. We derive $T_{12}$ for a gas of strings with a fixed set of quantum numbers as:

\eq{t12}{T_{12} = -c\frac{\mu_0}{\sqrt{g_s}|p_d|}\frac{e^{-4\omega}}{\alpha'^2}(w_1^2b_1^2 + w_2^2b_2^2).}  

\noindent Again, since $c$ is infinitesimal, we conclude that the above term contributes to (\ref{shapeq}) as:

\eq{shpeq}{\ddot c + (dH -\mathcal H_1 - \mathcal H_2)\dot c + c\frac{8\pi G_D}{2\alpha'}\frac{N\mu_0}{\sqrt{g_s}|p_d|} = 0,}  

\noindent where $N$ is some positive number which corresponds to the degeneracy of such a term once we sum over all massless modes. Again we see that this modulus is stabilized. Using (\ref{newtred}), and setting the scale of stringy physics to the Planck scale, we compute the mass of this modulus to be:

\eq{mshape}{m^2 \sim \frac{\mu_0}{a^3|p_d|},}    

\noindent which is similar to the masses found for all the other moduli fields (\ref{modmassa}). The case for extra shape moduli, such as what would appear for higher dimensional tori, should be no different, but nevertheless should be examined. We postpone this study to a future paper.

\par

We conclude by addressing a few points concerning the consistency of this framework alluded to in the introduction, and offering our closing thoughts. The first thing to check is that we have remained within the regime of low energy effective theory throughout the evolution of our system on the way to moduli stabilization. This is easily done by taking the trace of Einstein's equations, from which we find that:

\eq{tree}{R = \frac{-8\pi G_D}{D-1}T^\mu_\mu.}

\noindent Refering to (\ref{ee}) and (\ref{dc1}), we see that this implies that when the system is fluctuating around the stabilization point in moduli space (again using (\ref{newtred}), and setting the string scale to be near the Planck scale):

\eq{tree1}{R \sim \frac{\mu_0}{a^3|p_d|}(...),}

\noindent where the fluctuations of the moduli fields will appear in the parentheses. Clearly, since the prefactor is of the order of electron volts (so that we evade fifth force constraints-- see (\ref{modmassa})\cite{sp1}\cite{sp2}), one quite easily satisfies the condition (\ref{cr}):

\eq{tree2}{R \ll \frac{1}{\alpha'}.} 

\noindent We make the comment here that were we to consider {\it massive} string modes as the source terms for the dynamics, one typically could not satisfy (\ref{tree2}). 

\par

Another aspect of our model that should be checked, is that the energy momentum tensor is covariantly conserved:

\eq{covcons}{\nabla_\mu T^\mu_\nu = 0 ~\to~ \dot\rho + \sum_{\mu = 1}^D H_\mu(\rho + p_\mu) = 0.}

\noindent It is easy to check that this is (quite remarkably) satisfied as an identity by the coupled string gas-dilaton system (\ref{rho})-(\ref{pa}) and (\ref{rd})-(\ref{pd}). In taking the time derivative of $\rho$, which is a function of the metric and the dilaton, one finds that the time derivatives of the metric factors are cancelled by contributions from the terms proportional to the Hubble factors in the above, and the terms involving time derivatives of the dilaton are cancelled by the same, subject to the equations of motion for the dilaton field.  

\par

We also wish to note that since we have chosen a simple toroidal compactification for our compact manifold, one does not have any external sources for the ($B_{\mu\nu}$) flux fields, as would be the case if we were to have used an orbifold compactification (the orbifold fixed planes would then source the flux fields). Hence it is consistent for us to set these fields to be vanishing. A further consequence of this compactification scheme, is that a gas of D-strings had better contain an equal number of RR charges and anti-charges if Gauss's law is to be satisfied. Since we are considering low energy phenomena, any hydrodynamical fluid of such a collection of D-strings will look RR charge neutral, and will not source RR fields, which also allows us to set these fields to be zero. 

\par

As this concludes the body of our report, we feel that a summary would be in order at this point to realign our perspective before we offer our closing thoughts. We have demonstrated in this paper, that it is possible to stabilize all of the moduli in the context of a general toroidal compactification in the context of bosonic string theory. The workhorse of this stabilization mechanism is a gas of massless F-string and D-string modes. In the introduction, we provided several a priori arguments as to why one should focus exclusively on these modes, after which we derived the energy momentum tensor of a general string gas. We then showed how one quite naturally ends up with only the massless string states if we are at an energy that is even one order of magnitude below the Hagedorn temperature. Upon studying the dynamics of the various moduli fields coupled to this string gas, we found that we obtain stabilization of the radial moduli, given the stabilization of the dilaton. We then found that we obtained dilaton stabilization, given the stabilization of the radial moduli. However, upon examining the dynamics of the coupled radion-dilaton-string gas system, we found that a zero mode appears for one of the moduli. The physical origin of this zero mode was uncovered, and we then looked for ways of lifting this zero mode. We used the arguments presented in the introduction to focus on one possibility-- the introduction of a gas of D-strings. We found that at the point in moduli space where the D-string tension was equal to the F-string tension, stabilization of all moduli resulted. We then examined the question of how one would obtain a situation where the tensions of the two classes of strings were equal, and found that in a situation of broken supersymmetry (as would neccesarily be the case in the present epoch, especially if we had a gas of massless D-strings propagating through our space), that this was certainly possible consistent with weak coupling. We then considered the consistency of our framework with the low energy effective theory that provides us with our gravitational theory, and found that our model passes all of the tests we put to it (phenomenological or theoretical) quite easily. Hence, albeit in the context of a specific model, we feel that we have presented a model of late time cosmology that offers a solution to the moduli problem. We conlude with a discussion of the strengths and weaknesses of this approach, and open questions that this new mechanism begs.

\section{Conclusions and Outlook}

In the course of this study, some rather remarkable properties of massless string modes were uncovered. Not only did these properties prove crucial in stabilizing all of the moduli in our framework, but had, in addition, many other desirable phenomenological features associated with them. In general, we expect plenty of such strings to be present in the present epoch. As massless objects, these are likely to be copiously produced even if a preceding epoch of inflation took place. A very desirable feature of our framework is the distinct lack of fine tuning involved. Unlike most other attempts at moduli stabilization, which require fluxes to be turned on in a certain manner, with specific brane configurations to be present, introducing a string gas allows no such freedom. From the form of the coupling of the string gas to gravity (\ref{mod}), we see that we do not need to tune the minima of our 'moduli potentials' to avoid an unacceptably large De-Sitter phase of evolution. Another advantage is that evading fifth-force constraints without overclosing the universe is easily accomplished by a string gas, again without any fine tuning (see (\ref{modmassa}), (\ref{cosmomass}), \cite{sp1} and \cite{sp2}). In \cite{sp3}, we will investigate further the observational consequences of having an all pervasive gas of strings stabilizing extra dimensions at the present epoch, which we feel might be within the reach of experiments in the near future. However all of these results are just a desirable aside to the main result, that we can consistently stabilize all of the moduli fields in our compactification within our framework. 
\par
It is prudent at the very least, to address the shortcomings of such an approach, not only to inform future avenues of investigation, but to also contrast to other approaches to moduli stabilization (see \cite{eva} and references therein). Clearly, one cannot hope to stabilize moduli fields associated with the compactifications of string theory solely using string gases if the the compact manifold that results does not admit topological 1-cycles (see however \cite{nyc4}). That is, since 1-dimensional extended objects need 1-cycles to admit winding quantum numbers, manifolds that do not admit these (such as Calabi-Yau spaces) will not readily admit to our mechanism. In addition, we have not yet presented any set-up that generates an inflationary epoch for our non-compact spacetime within our moduli stabilization mechanism (however there are promising indications one such a mechanism exists \cite{sp4}, see also \cite{infl}). We presume that this situation will be markedly different if our framework was generalized to include brane gases. For in this case, not only can we hope to stabilize compact sub-manifolds that do not admit 1-cycles (but do admit higher dimensional cycles) through the use of higher dimensional branes, but one might also expect inflation to fall out of such a set-up. One can envisage that the various inter-brane potentials that arise might generate an inflationary epoch on the way to moduili stabilization. However in this case, the framework we propose starts to look a lot more like the more standard approaches to moduli stabilization \cite{eva}, and it is highly likely that these approaches will overlap significantly. It is our belief that the mechanism that will eventually succeed as the main candidate for stabilizing all of the moduli fields of string theory, will certainly be some combination of the flux-compactification and the string/brane gas approach. However, insofar as providing a self contained, consistent and natural mechanism for moduli stabilization in the context we have chosen to study, we feel that the string gas approach has several comparative advantages over the flux compactification approach. It is our hope that, at the very least, the results uncovered in this report will help motivate the further study of string gas cosmology, not only as a mechanism to stabilize moduli, but also as a viable, alternative stringy cosmology.    

\section{Acknowledgements}

We wish to thank Joseph Conlon, Neil Turok, Sanjaye Ramgoolam, Thorsten Battefeld, Hoartiu Nastase, Tirthabir Biswas and Keshav Dasgupta for useful discussions and criticisms. We also wish to thank Robert Brandenberger for his continued encouragement, comments on the manuscript and many valuable discussions. Special thanks to Anne Davis and Bill Spence for providing a once-in-a-PhD opportunity to spend some time at two of the U.K.'s best research institutions. This work is supported at McGill by an NSERC Discovery grant and funds from McGill University.

\appendix

\section{The Energy Momentum Tensor}

We offer in this appendix, a derivation of the string gas energy momentum tensor based on micro-physical considerations. That is, we derive the space-time energy momentum tensor of a single string in a given background, and proceed to hydrodynamically average to obtain the result for a gas of such strings. Before we do this, we should comment on the fact that we are going to be using aspects of the string spectrum in a background that is varying in time. Certainly, string theory in a time-dependent background is an active and current area of research, in which many unresolved issues remain. However, we are going to be doing string theory in which the background varies on a time scale that is many, many orders of magnitude slower than the string scale, and as such one would expect to inherit many features of the string spectrum in this very weakly time dependent setting. This intuition is justified rather more rigorously in \cite{sp1}, where we discuss quantization of the string in such a time dependent background and show that (in what is essentially an application of the adiabatic theorem of quantum mechanics), we have the same spectrum of states in a weakly time dependent background, with the energies now very slowly varying (compared to the string scale) functions of time. We arrive at this conclusion from considering how the constraint algebra of string theory changes in the context of a weakly time-dependent metric.  

\par

Returning to the problem at hand, we begin with the Nambu-Goto action:

\eq{sng}{S = -\frac{1}{2\pi\alpha'}\int d^2\sigma \sqrt{-h},}

\noindent where the induced worldsheet metric $h_{ab}$ is given by:

\eq{wsm}{h_{ab} = g_{\mu\nu}(X)\partial_aX^\mu\partial_bX^\nu,}

\noindent and the spacetime metric (in general a function of the worldsheet fields $X^\mu$), is assumed to have the form:

\begin{eqnarray}
\label{goo}
g_{00} &=& g_{00}(X^0)\\
\label{gij}
g_{ij} &=& \delta_{ij}a^2(X^0)\\
\label{gab}
g_{ab} &=& \gamma_{ab}(X^0). 
\end{eqnarray}

\noindent By varying (\ref{sng}) with respect to the background metric, one obtains the {\it spacetime} energy momentum tensor as:

\eq{stemt}{T^{\mu\nu} = \frac{2}{\sqrt{-g}}\frac{\delta S}{\delta g_{\mu\nu}}.}

\noindent An arbitrary variation of the background metric induces a variation of the worldsheet metric as:

\eq{relvar}{\delta^{\lambda\beta}g_{\mu\nu} = \delta^{\lambda}_{\mu} \delta^{\beta}_{\nu}\delta^{D+1}(X^\tau - y^\tau)~~\to~~\delta^{\lambda\beta} h_{ab} = \partial_aX^\lambda\partial_bX^\beta\delta^{D+1}(X^\tau - y^\tau),}

\noindent where the unmatched indices $\lambda$ and $\beta$ imply that we perturb only these components of the metric tensor, and $\delta^{D+1}(X^\tau - y^\tau)$ is a delta function in $D+1$ spacetime dimensions. The result of this variation yields an expression for the energy-momentum tensor for a single string as:

\eq{emt1s}{T^{\mu\nu} = \frac{-1}{\sqrt{-g}2\pi\alpha'}\int d^2\sigma \sqrt{-h}h^{ab}\partial_aX^\mu\partial_bX^\nu \delta^{D+1}(X^\tau - y^\tau).}

\noindent Before we can proceed, we must pick a gauge to work in. We choose to work in conformal gauge, defined by:

\eq{cg}{h_{ab}= \lambda \begin{pmatrix} -1&0\\0&1\end{pmatrix},}

\noindent where we keep this up to an arbitrary positive factor, which we will choose further on. From (\ref{wsm}), we see that this gauge choice implies the conditions:

\eq{wcc1}{g_{\mu\nu}\dot X^\mu\dot X^\nu + g_{\mu\nu}X'^\mu X'^\nu = 0}
\eq{wcc2}{g_{\mu\nu}\dot X^\mu X'^\nu = 0.}

\noindent It can be shown that even though we are in a (weakly) time-dependent background (in particular one that is not flat), one can still make this gauge choice simultaneous with the condition:

\eq{tdo}{X'^0 = 0,}

\noindent where the prime denotes a derivative with respect to the spacelike worldsheet co-ordinate. These condition will prove very useful in what follows.
\par
Upon examining (\ref{emt1s}), we see that in order to make use of the spacetime delta functions, one would have to use the following transformation:

\eq{jaco}{d^2\sigma = \frac{dX^0 dX^a}{|\dot X^0||X'^a|},}

\noindent where $X^a$ is the string co-ordinate field along any wound compact direction (which one we pick is insignificant). The term in the denominator comes from evaluating the jacobian of this transformation, subject to the condition (\ref{tdo}). Note that we picked the particular co-ordinates $X^0$, because it is a monotonic function in time, and $X^a$ because it is easy to evaluate for wound strings. Using the constraints (\ref{wcc1}) and (\ref{wcc2}), we see that:

\eq{wcc3}{\dot X^0 = \frac{2\pi\alpha'}{\sqrt{-g_{00}}}\sqrt{g^{ij}P_iP_j + \frac{1}{(2\pi\alpha')^2}g_{ij}X'^iX'^j}~,}

\noindent where we use the fact that in the gauge we're in:

\eq{mom}{P_\mu = \frac{g_{\mu\nu}\dot X^\nu}{2\pi\alpha'}.}

\noindent The expression in the sqaure root in (\ref{wcc3}) is given by the $L_0$ constraint in our constraint algebra \cite{pol} (see also (\ref{energy}) and (\ref{lmc})), and is equal to the energy of the closed string:

\eq{energya}{\epsilon = \sqrt{|p_{d}|^2 + (n,\gamma^{-1} n) + \frac{1}{\alpha'^2}(w,\gamma w) + \frac{1}{\alpha'}[2(n,w) + 4(N-1)]},}

\noindent where the worldsheet zero modes give us the contributions $|p_d|$ for momentum along the non-compact directions, as well as the terms containing the winding and momentum quantum numbers for the compact dimensions. All the other fourier modes give us the oscillator contributions. We write the above as:

\eq{wcc4}{\dot X^0 = \frac{2\pi\alpha'}{\sqrt{-g_{00}}}\epsilon.} 
 
\noindent As for the second factor (\ref{jaco}) coming from the Jacobian, we see that for any string wound $w$ times around the $a^{th}$ direction:

\eq{wcc5}{X'^a = w.}

\noindent Recall that we have to sum over all zeroes of the delta function along the $a^{th}$ direction when evaluating the integral (\ref{emt1s}) after our change of variable (\ref{jaco}). Hence the contribution (\ref{wcc4}) is cancelled by the fact that when a string winds $w$ times around a particular direction, the argument of the delta function is zero precisely $w$ times. Hence we evaluate (\ref{emt1s}) using the gauge fixing conditions and the results just obtained (\ref{wcc3}) - (\ref{wcc5}) as:
 
\begin{eqnarray}
T^0_0 &=& \frac{\epsilon}{\sqrt{g_s}}\delta^{D-1}(X^\tau - y^\tau),\\
T^i_i &=& \frac{p^ip_i}{\epsilon\sqrt{g_s}}\delta^{D-1}(X^\tau - y^\tau),\\
T^a_a &=& \frac{1}{\epsilon\sqrt{g_s}}\Bigl(\frac{n_a^2}{b_a^2} - \frac{w_a^2b_a^2}{\alpha'^2}\Bigr)\delta^{D-1}(X^\tau - y^\tau),
\end{eqnarray}

\noindent where $g_s$ is the determinant of the spatial part of the metric, and $\epsilon$ defined by (\ref{energya}). We have deliberately only indicated the diagonal elements as they are the only elements which will survive a hydrodynamical averaging. We do this averaging as follows: keeping the quantum numbers $p_d^2, \vec w, \vec n$ and $N$ fixed, we sum the contributions over a distribution of such strings, where the momentum along the non-compact directions is distributed isotropically. Doing so gives us the result (\ref{rho})-(\ref{pa}):

\begin{eqnarray}
\label{rho3}
\rho &=& \frac{\mu_0\epsilon}{\sqrt{{g}_s}},\\
\label{pi3}
p^i &=& \frac{\mu_0}{\sqrt{{g}_s}\epsilon}|p_{d}|^2/d,\\
\label{pa3}
p^a &=& \frac{\mu_0}{\sqrt{{g}_s}\epsilon}[\frac{n_a^2}{{b}_a^2} - \frac{w_a^2 b_a^2}{\alpha'^2}].
\end{eqnarray}

\noindent This is exactly the same result one would obtain if one were to introduce the following action for the string gas:

\eq{sgact}{S = -\int d^{D+1}x\sqrt{-g_{00}}\mu_0\epsilon~.}  

\noindent Note that in the above, we only assumed that the metric tensor was diagonal. Hence our result is valid even if we to begin in a frame such that $g_{00}$ was a function of time (rather than equal to $-1$), such as what would be required if we were to transform from the string frame to the Einstein frame using (\ref{ct}), where we would like to end up with an Einstein frame metric of the form:

\eq{efm}{g_{\mu\nu} = diag[-1,a^2(t),a^2(t),a^2(t),b_1^2(t),...b_p^2(t)].}    

\noindent Hence our energy momentum tensor in the Einstein frame would simply be the transformation of the above result using (\ref{ct}). As for the issue of the energy momentum tensor of a D-string gas, it is straightforward to check that if we are in a regime where:

\eq{dscond}{\frac{\partial \Phi(X^0)}{\partial X^\mu} \ll \frac{1}{\sqrt{\alpha'}},}  

\noindent then the only difference in the steps leading to the above derivation would be terms proportional to derivatives of $\Phi$ with respect to the fields $X^\mu$. Since these will appear in conjunction with terms involving worldsheet derivatives of the fields $X^\mu$, which will be of the string scale, the above condition will ensure that they will be negligable. For example if we demand that $\partial_{X^0}\Phi < 10^{-6}/\sqrt{\alpha'}$, then almost certainly such terms will be insignificant and the net effect of having the extra dilaton coupling:

\eq{dactor}{S = \frac{-T_1}{2\pi}\int d^2\sigma e^{-\Phi}\sqrt{-h},}    
\noindent will be the replacement of $\alpha'$ with $\hat\alpha = e^{\Phi}/T_1$ in (\ref{rho3})-(\ref{pa3}). This is essentially a generalization of the argument given in the appendix of \cite{sp1} where the weak time dependence of the metric was shown to give a spectrum (\ref{energya}) that evolved adiabatically (where the metric factors were allowed to depend on time).

\section{Summation Over Massless Modes}

Consider the expression for the energy of a closed string (\ref{tenergy}) at the self dual radius, with the dilaton at its expectation value ($\gamma_{ab} = \alpha'\delta_{ab}$, $\omega = 0$):

\begin{equation}  
\label{msqua}
m^2 = \frac{1}{\alpha'}\Bigl[(n+w,n+w) + 4(N-1)\Bigr].
\end{equation}

\noindent We can read off that from this that the massless states are those which simultaneously satisfy the following set of equations:

\begin{eqnarray}
\label{zero}
(n+w,n+w) &=& 4(1-N),\\
\label{lc}
N + (n,w) &\geq& 0,
\end{eqnarray}

\noindent where the last equation is the level matching constraint. The quantum numbers that satisfy these conditions are tabulated below:

\begin{center}
\begin{tabular}{c|ccc|}
$N~$&$(n,n)~$&$(w,w)~$&$(n,w)~$\\
\hline
1&0&0&0\\
1&1&1&-1\\ 
0&1&1&1\\
0&2&2&0\\
\hline
0&1&3&0\\
0&3&1&0\\
0&4&0&0\\
0&0&4&0\\
\end{tabular}
\end{center}

\noindent We note that whether or not the particular modes indicated above are realised or not depends on the number of extra dimensions. For instance for one or two extra dimensions, it will not be possible to satisfy $(w,w) = 3$ etc. We shall be most interested in the first four possibilities, classed separately above. This is because these states are not only massless at the self dual radius, with $\omega = 0$, they are also massless to first order away from this point in moduli space. The first possibility ($N=1,n=w=0$) corresponds to unwound gravitons. As mentioned before, these states do not directly contribute to the dynamics of the moduli fields, as can be seen from (\ref{dc}) and (\ref{eld}). The second possibility, in light of the level matching constraint can be seen to imply states of the form $N=1$, $n=\pm w$, $(w,w)=1$, corresponding to singly wound strings. These strings have an equal and opposite quantum of momentum along the same dimension, at oscillator level 1. The third possibility (again re-applying the level matching constraint) yields $N=0$, $n=w$, $(w,w)=1$ as massless quantum numbers corresponds to a singly wound state at oscillator level zero, with one quantum of momentum of the same sign along the same dimension. The fourth possibillity ($N=0$, $(n,n) = (w,w) = 2$, $(n,w) = 0$) will also prove to be relevant. The second class of states are all at oscillator level zero, and correspond to various multiply wound/unwound strings with/without motion along various cycles of our torus. We now examine the behaviour of states with the quantum numbers tabulated above, as we move away from the rather special point in moduli space that we began in. To do this, we perturb our moduli as follows:

\begin{eqnarray*}
\label{gp}
&\tilde{\gamma}& = I - \Delta,\\
&\tilde{\gamma}^{-1}& = I + \Delta~~;~~||\Delta|| \ll 1,\\
&\omega& = \eta~~;~~ \eta \ll 1,
\end{eqnarray*} 

\noindent where $\Delta$ is has a matrix norm that is much less than 1. Expanding the formula for the mass of a closed string (\ref{tenergy}), we see that:

\eq{variss}{\alpha' \delta m^2 = (n,\Delta n) - (w,\Delta w) -4\eta[(w,w) + (n,w) + 2(N-1)],}

\noindent from which we see that only the first class outlined above  remain massless to first order. For the first three states in the above table, the contribution from the metric fluctuations vanishes because $\vec n$ and $\vec w$ are equal to each other up to a sign. For the states with quantum numbers given by $N=0$, $(n,n) = (w,w) = 2$, $(n,w) = 0$, only those states that have entries in the same rows have vanishing fluctautions. That is, the condition $(n,w)=0$ implies that these states either have no entries in common, or two entries in common. The contribution from the metric fluctations would be non-vanishing in the former case, but for the latter, states of the form below remain massless to first order:

\begin{eqnarray*}
w &=& (1,1)~ n = \pm(1,-1),\\
w &=& (1,-1)~ n = \pm(1,1),\\
w &=& (-1,1)~ n = \pm(1,1),\\
w &=& (-1,-1) ~ n = \pm(1,-1)
\end{eqnarray*}

\noindent where we only indicate the entries which the two vectors have in common. Our problem is now reduced to the following: having determined which states enter the summations in (\ref{dc}) and (\ref{eld}), we have to compute their contribution to the driving term taking care to account for all of them. We shall only compute the driving term for the radial moduli, as the case for the dilaton proceeds identically. Recalling that the unwound graviton states ($N=1$,$n=w=0$) do not contribute to the driving term, we compute the contribution of the states with quantum numbers $N=1$, $n=\pm w$, $(w,w)=1$ to the driving term of (\ref{dc}) as:

\begin{equation}
\label{mt}
2\times\frac{8\pi G_D \mu_{0}}{\alpha'^{1/2}\sqrt{g_s}}\frac{1}{|p_d|}\Bigl[ \frac{1}{b_a^2} - \frac{b_a^2e^{-4\omega}}{\alpha'^2} + \frac{2}{D-1}(\sum_{c=1}^{p} \frac{b_c^2e^{-4\omega}}{\alpha'^2} - \frac{pe^{-2\omega}}{\alpha'}) \Bigr].
\end{equation}     

\noindent The factor of $2$ comes from the overall degeneracy of the states which can appear either as, for example $w = (0,0,0...1,0...)$, $n = (0,0,0...-1,0...)$ or with the opposite signs, with the factor $|p_d|$ coming from the factor of the energy in the denominator in the summand in (\ref{dc}). The states with quantum numbers $N=0$, $n=w$, $(w,w)=1$ sum to yield an identical contribution to the driving term. It is non-trivial, but straightforward to compute the contribution of the states corresponding to $N=0$, $(n,n) = (w,w) = 2$, $(n,w) = 0$, where we find that we end up with the identical contribution as above, except now with the prefactor of $8(p-1)$, where $p-1$ has the combinatorial origin as being the number of ways one can pick two entries to be identical out of $p$ choices, and $8$ corresponds to the overall degeneracy of these states (as indicated above). Hence the overall driving term is

\begin{equation}
\label{mt3}
(8p-4)\times\frac{\pi G_D \mu_{0}}{\alpha'^{1/2}\sqrt{g_s}}\frac{2}{|p_d|}\Bigl[ \frac{1}{b_a^2} - \frac{b_a^2e^{-4\omega}}{\alpha'^2} + \frac{2}{D-1}(\sum_{c=1}^{p} \frac{b_c^2e^{-4\omega}}{\alpha'^2} - \frac{pe^{-2\omega}}{\alpha'})\Bigr],
\end{equation}     

\noindent where expanding around $b_a = \sqrt{\alpha'}$, $\omega = 0$ gives us (\ref{dc1}). The evaluation of the driving term for the dilaton proceeds analogously, as does the case for the summation over massless D-string modes.

\end{document}